\documentclass{aa}  

\usepackage{graphicx}
\usepackage{txfonts}
\usepackage{natbib}
\bibpunct{(}{)}{;}{a}{}{,} 

\usepackage{color}

\usepackage{hyperref}
\hypersetup{breaklinks=True,
    colorlinks=true,
    linkcolor=blue,
    citecolor=blue,
    filecolor=cyan,      
    urlcolor=blue,
    pdftitle={Atomic and molecular gas properties during cloud formation},
    pdfauthor={J. Syed}}

\begin{document}

   \title{Atomic and molecular gas properties during cloud formation}

   \subtitle{}

   \author{J. Syed\inst{1}
   \and Y. Wang\inst{1}
   \and H. Beuther\inst{1}
   \and J. D. Soler\inst{1}
   \and M. R. Rugel\inst{2}
   \and J. Ott\inst{3}
   \and A. Brunthaler\inst{2}
   \and J. Kerp\inst{4}
   \and M. Heyer\inst{5}
   \and R. S. Klessen\inst{6,7}
   \and Th. Henning\inst{1}
   \and S. C. O. Glover\inst{6}
   \and P. F. Goldsmith\inst{8}
   \and H. Linz\inst{1}
   \and J. S. Urquhart\inst{9}
   \and S. E. Ragan\inst{10}
   \and K. G. Johnston\inst{11}
   \and F. Bigiel\inst{4}
          }

   \institute{Max-Planck-Institut f\"ur Astronomie, K\"onigstuhl 17, 69117 Heidelberg, Germany\\
              \email{syed@mpia.de}
              \and Max-Planck-Institut f\"ur Radioastronomie, Auf dem H\"ugel 69, 53121 Bonn, Germany
              \and National Radio Astronomy Observatory, PO Box O, 1003 Lopezville Road, Socorro, NM 87801, USA
              \and Argelander-Institut für Astronomie, Auf dem H\"ugel 71, 53121 Bonn, Germany
              \and Department of Astronomy, University of Massachusetts, Amherst, MA01003, USA
              \and Universit\"at Heidelberg, Zentrum für Astronomie, Institut f\"ur Theoretische Astrophysik, Albert-Ueberle-Str. 2, 69120 Heidelberg, Germany
              \and Universit\"at Heidelberg, Interdisziplin\"ares Zentrum für Wissenschaftliches Rechnen, INF 205, 69120 Heidelberg, Germany
              \and Jet Propulsion Laboratory, California Institute of Technology, 4800 Oak Grove Drive, Pasadena, CA 91109, USA
              \and Centre for Astrophysics and Planetary Science, University of Kent, Canterbury CT2 7NH, UK
              \and School of Physics and Astronomy, Cardiff University, Queen’s Buildings, The Parade, Cardiff CF24 3AA, UK
              \and School of Physics and Astronomy, E.C. Stoner Building, The University of Leeds, Leeds LS2 9JT, UK
             }

   \date{Received XX May XXXX; accepted 12 August 2020}

 
  \abstract
   {Molecular clouds, which harbor the birthplaces of stars, form out of the atomic phase of the interstellar medium (ISM). To understand this transition process, it is crucial to investigate the spatial and kinematic relationships between atomic and molecular gas.}
   {We aim to characterize the atomic and molecular phases of the ISM and set their physical properties into the context of cloud formation processes.}
   {We studied the cold neutral medium (CNM) by means of \ion{H}{i} self-absorption (HISA) toward the giant molecular filament GMF20.0-17.9 (distance=$3.5\rm\,kpc$, length $\sim$170$\rm\,pc$) and compared our results with molecular gas traced by \element[][13]{CO} emission. We fitted baselines of HISA features to \ion{H}{i} emission spectra using first and second order polynomial functions.}
   {The CNM identified by this method spatially correlates with the morphology of the molecular gas toward the western region. However, no spatial correlation between HISA and \element[][13]{CO} is evident toward the eastern part of the filament. The distribution of HISA peak velocities and line widths agrees well with \element[][13]{CO} within the whole filament. The column densities of the CNM probed by HISA are on the order of $10^{20}\rm\,cm^{-2}$ while those of molecular hydrogen traced by \element[ ][13]{CO} are an order of magnitude higher. The column density probability density functions (N-PDFs) of HISA (CNM) and \ion{H}{i} emission (tracing both the CNM and the warm neutral medium, WNM) have a log-normal shape for all parts of the filament, indicative of turbulent motions as the main driver for these structures. The $\rm H_2$ N-PDFs show a broad log-normal distribution with a power-law tail suggesting the onset of gravitational contraction. The saturation of \ion{H}{i} column density is observed at $\sim$25$\rm\,M_{\odot}\,pc^{-2}$.}
   {We conjecture that different evolutionary stages are evident within the filament. In the eastern region, we witness the onset of molecular cloud formation out of the atomic gas reservoir while the western part is more evolved, as it reveals pronounced $\rm H_2$ column density peaks and signs of active star formation.}


   \maketitle
%
\section{Introduction}

    Molecular clouds play a key role in star formation processes. Stars are born in the dense interiors of molecular clouds that form out of the atomic phase of the highly turbulent interstellar medium (ISM) \citep{1981MNRAS.194..809L,2012MNRAS.424.2599C,2014prpl.conf....3D,2014ApJ...790...10S,2016SAAS...43...85K}. Molecular clouds consist mainly of molecular hydrogen \citep{2003RPPh...66.1651L,2004RvMP...76..125M,2007ARA&A..45..565M,2014prpl.conf....3D}. However, the cloud formation process out of the diffuse atomic phase is still not well constrained. According to the standard photodissociation region (PDR) model, layers of cold atomic hydrogen can effectively shield the cloud from photo-dissociating UV radiation at sufficiently high densities, allowing a more complete conversion of \ion{H}{i} to its molecular form. The cold neutral medium (CNM) with temperatures of $\leq 300\rm\,K$ and volume densities of $10-100\rm\,cm^{-3}$ \citep{1977ApJ...218..148M,2003ApJ...586.1067H,2003ApJ...587..278W,2009ARA&A..47...27K} is thought, due to its relatively high density, to be a key component in the conversion process from diffuse atomic hydrogen to its molecular phase. Constraining the physical and dynamical properties of the CNM is therefore crucial to understand early cloud formation processes.

    The CNM is a major constituent of the ISM \citep[see e.g.,][]{2001RvMP...73.1031F,2003ApJ...586.1067H}. Even though the observation of the \ion{H}{i} 21cm line allows one to study the properties of atomic hydrogen in general, it is difficult to attribute certain properties to different components of \ion{H}{i}. In pressure equilibrium, atomic hydrogen can exist in different phases \citep[e.g.,][]{1977ApJ...218..148M,2003ApJ...587..278W}. Observations of \ion{H}{i} 21cm line emission are generally attributed to both warm neutral medium (WNM) and CNM. To separate the WNM from the CNM, we make use of the presence of \ion{H}{i} self-absorption \citep[HISA; see e.g.,][]{1972A&A....18...55R,1974AJ.....79..527K,1988A&A...201..311V,1993A&A...276..531F,2000ApJ...540..851G,2005ApJ...626..195G,2005ApJ...626..214G,2003ApJ...598.1048K,2018MNRAS.479.1465D,2020A&A...634A.139W} to trace the cold atomic phase. \ion{H}{i} self-absorption is found throughout the Milky Way in various environments. Many studies have focused on the detection of HISA, first detected in 1954 \citep{1954AJ.....59..324H,1955ApJ...121..569H}, in known sources, but statistical treatments of the kinematic properties and densities of the CNM in large-scale high-resolution maps are still rare. 
    
    For HISA to be detected, sufficient background emission of warmer gas along the line of sight is required. Since the warm component of atomic hydrogen is more diffuse, it fills up a larger volume than the cold component \citep{1977ApJ...218..148M,2005fost.book.....S,2009ARA&A..47...27K}.
    \ion{H}{i} self-absorption occurs when a cold \ion{H}{i} cloud is located in front of a warmer \ion{H}{i} emitting cloud. Self-absorption can occur within the same cloud but can also be induced by an emitting cloud in the far background that has the same velocity as the absorbing medium with respect to the local standard of rest $v_{\rm LSR}$. Therefore, the clouds do not have to be spatially associated for HISA to be observable.
    While absorption against strong continuum sources does yield a direct measurement of the optical depth, the discreteness of the sources only delivers an incomplete grid of optical depth measurements \citep[e.g.,][]{2020A&A...634A..83W}. The interpolation of optical depths across an entire \ion{H}{i} cloud is challenging. Therefore, the great advantage of HISA is that larger areas of cold atomic hydrogen can be mapped.
    
    Large filamentary gas structures, also known as Giant Molecular Filaments (GMFs), are suitable to study the CNM on large scales. These objects are the largest coherent structures found in the Milky Way and are subject of many studies probing the physical properties of the Galactic ISM \citep{2010ApJ...719L.185J,2014ApJ...797...53G,2014A&A...568A..73R,2015ApJ...815...23Z,2018ApJ...864..153Z,2016A&A...590A.131A}.
    We study the hydrogen content by means of HISA, atomic and molecular line emission toward the giant molecular filament GMF20.0-17.9 \citep{2014A&A...568A..73R}. We address the physical processes driving the kinematics of the CNM and the properties that lead to molecular cloud formation.
    
\begin{figure*}[!htbp]
      \centering
        \includegraphics[width=1.0\textwidth]{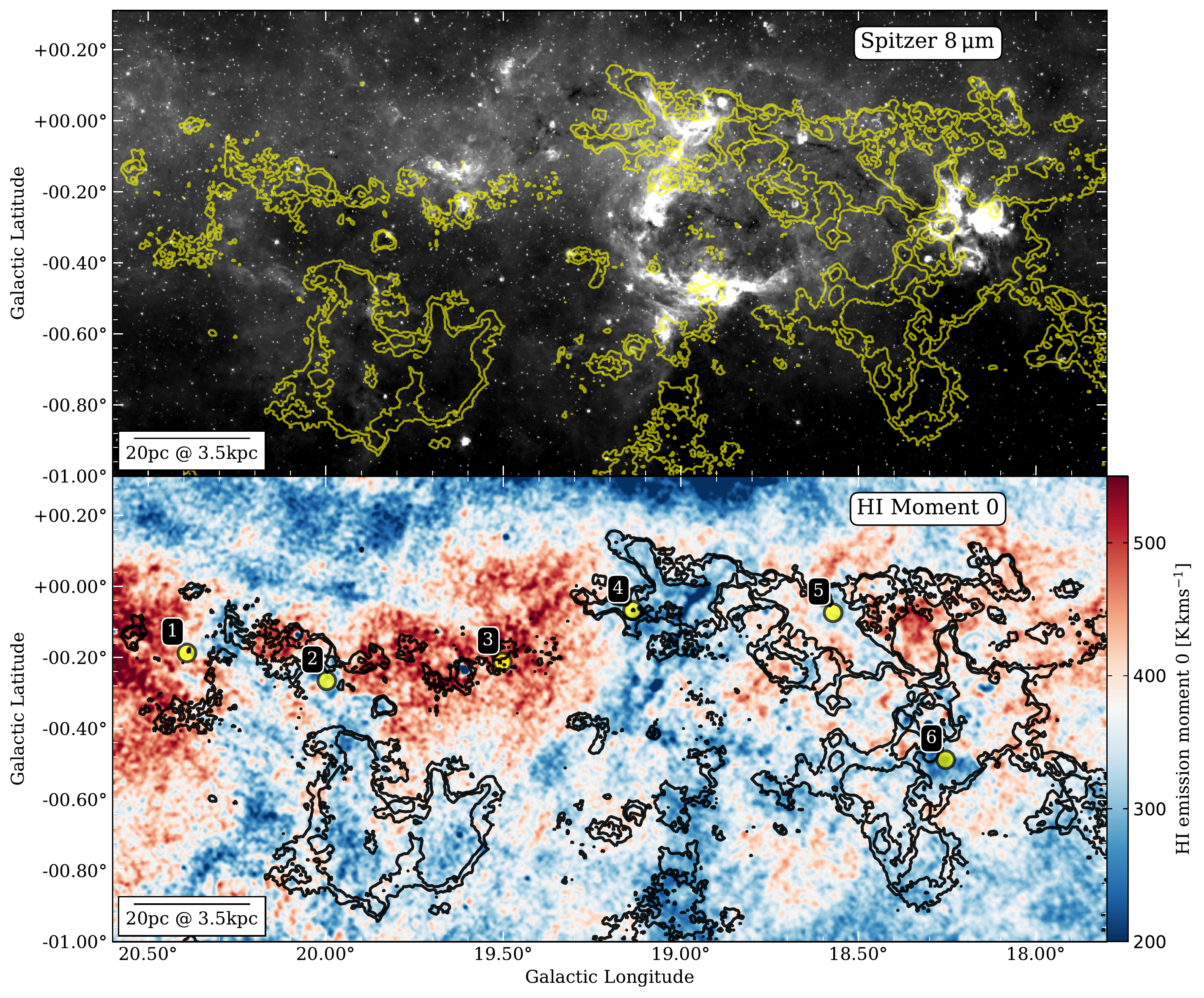}
      \caption[]{GMF20.0-17.9 overview. \textit{Top panel:} Spitzer GLIMPSE $8\rm\,\mu m$ image of GMF20.0-17.9 \citep{2009PASP..121..213C}. The color scale is chosen to bring IRDC features to prominence. \textit{Bottom panel:} \ion{H}{i} integrated emission for a small velocity interval from 44.5--$47.5\rm\,km\,s^{-1}$. The yellow and black contours show the integrated \element[ ][13]{CO} emission from $42$ to $57\,\rm km\,s^{-1}$ at the levels of $10.5\,\rm K\,km\,s^{-1}$ and $15\,\rm K\,km\,s^{-1}$, respectively. The yellow circles in the bottom panel mark the regions of the spectra shown in Fig.~\ref{fig:baseline_fits}.}
      \label{fig:HI_moment0_with_regions}
    \end{figure*}

     GMF20.0-17.9 was already identified in part by \citet{2013A&A...550A.116T}. Furthermore, \citet{2015ApJ...815...23Z,2018ApJ...864..153Z} define a subsection of this filament as a ``bone'' of the Scutum-Centaurus (SC) spiral arm. GMF20.0-17.9 is characterized by grouping several infrared dark clouds (IRDCs) into a single structure that is velocity-coherent as traced by \element[ ][13]{CO} emission. Figure~\ref{fig:HI_moment0_with_regions} shows an overview of GMF20.0-17.9. Prominent IRDC features along the \element[][13]{CO} emission are visible in the Spitzer $8\rm\,\mu m$ image, in particular toward the western part of the filament. It furthermore shows features of stellar activity. GMF20.0-17.9 extends from $20.2^{\circ}$ to $17.6^{\circ}$ in Galactic longitude and \mbox{$+0.3^{\circ}$} to \mbox{$-0.7^{\circ}$} in Galactic latitude. At the computed kinematic near distance of 3.3--$3.7\rm\,kpc$, this corresponds to a projected length of $\sim$170$\rm\,pc$. \citet{2014A&A...568A..73R} associate the velocity range of $37-50\rm\,km\,s^{-1}$ with GMF20.0-17.9. The filament is near the midplane of the Galaxy, and the velocity of the lower longitude part at $\sim$18$^{\circ}$ agrees fairly well with that of the near SC spiral arm \citep{2008AJ....135.1301V,2014ApJ...783..130R,2019ApJ...885..131R}. However, the sense of the velocity gradient of GMF20.0-17.9 as defined by \citet{2014A&A...568A..73R} goes against the trend of the spiral arm structure. \citet{2015ApJ...815...23Z} argue that the bone at $19.2^{\circ}\gtrsim\ell\gtrsim 18.6^{\circ}$, $b\approx -0.1^{\circ}$ traces the spine of the SC spiral arm well. This discrepancy is attributed to the different methodology for defining filaments and can be brought into agreement if only the lower longitude section of GMF20.0-17.9 is considered.
     
     The ATLASGAL survey \citep{2009A&A...504..415S} reveals several high-density clumps within GMF20.0-17.9, particularly in the western part of the filament. \citet{2019A&A...622A..52Z} identified young stellar object (YSO) populations within all currently known GMFs and derive a star formation rate (SFR) of $\mathrm{SFR}=1.2\cdot10^3\rm\,M_{\sun}\,Myr^{-1}$ and efficiency (SFE) of $\mathrm{SFE}=0.01$ for GMF20.0-17.9, which is consistent with SFEs of nearby star-forming regions \citep[see][and references therein]{2019A&A...622A..52Z}.

\section{Observations and methods}

\subsection{\ion{H}{i} 21 cm line and continuum}\label{sec:methods_and_observation}

    The following analysis employed the \ion{H}{i} and 1.4 GHz continuum data from the THOR survey \citep[The \ion{H}{i}/OH Recombination line survey of the inner Milky Way;][]{2016A&A...595A..32B,2020A&A...634A..83W}. The \ion{H}{i} and 1.4$\rm\,GHz$ continuum data include observations from the Karl G. Jansky Very Large Array (VLA) in both C- and D-configuration as well as single-dish observations from the Green Bank Telescope (GBT) and Effelsberg, respectively, to recover missing flux on short $uv$ spacings. Depending on the purpose of the analysis, different data products were utilized. For the analysis of \ion{H}{i} emission and the subsequent identification of HISA features, the combined THOR \ion{H}{i} data (VLA C+D + GBT) without continuum were used. The final data have been smoothed to an angular resolution of $\Delta\Theta=40\arcsec$ for better brightness sensitivity that is required especially for studying HISA. The rms noise in emission-free channels is $\sim$5$\rm\,K$. The spectral resolution is $\Delta v=1.5\rm\,km\,s^{-1}$. The final THOR 1.4 GHz continuum emission data (VLA C+D + Effelsberg) have an angular resolution of $\Delta\Theta=25\arcsec$.
    
    Additionally, optical depths were derived from \ion{H}{i} absorption against strong continuum sources. For that purpose, THOR-only data that comprise \ion{H}{i} emission with continuum were used. THOR C-array-only data have a higher angular resolution, making them suitable to study absorption against discrete continuum sources. Since these data consist of observations from the VLA in C-array configuration only, large-scale \ion{H}{i} emission is effectively filtered out. The THOR-only data have an angular resolution of $\Delta\Theta\sim 20\arcsec$, depending slightly on Galactic longitude. For more details about the THOR data, we refer to the two data release papers by \citet{2016A&A...595A..32B} and \citet{2020A&A...634A..83W}.
    
    We used the Galactic Ring Survey \element[ ][13]{CO}(1--0) data \citep[GRS;][]{2006ApJS..163..145J} to investigate the kinematic properties of the molecular gas and estimate the \element[ ][13]{CO} and $\rm H_2$ column density. The GRS \element[ ][13]{CO} data have an angular and spectral resolution of $\Delta\Theta=46\arcsec$ and $\Delta v=0.21\rm\,km\,s^{-1}$, respectively.

\subsection{\ion{H}{i} self-absorption (HISA) extraction}\label{sec:poly_fitting}

    In the following section, different methods to identify and extract HISA spectra from the \ion{H}{i} emission are discussed. Several approaches have been tested as the accurate extraction of HISA spectra poses a challenging task.
    
\begin{figure}[!htbp]
        \centering
            \resizebox{\hsize}{!}{\includegraphics{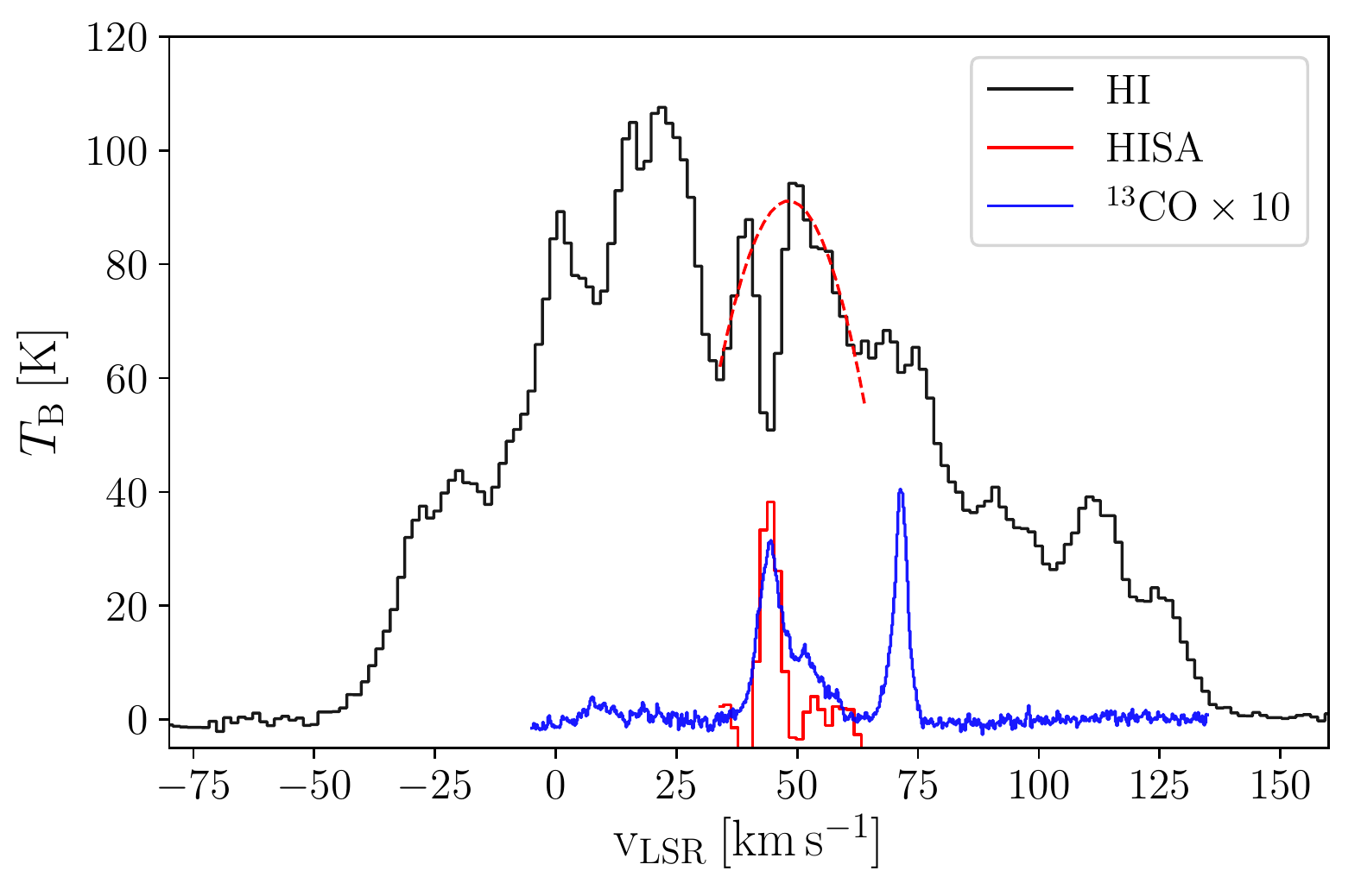}}
        \caption{\ion{H}{i}, HISA, and \element[][13]{CO} spectra. The black curve shows an example spectrum of \ion{H}{i} emission ($T_{\mathrm{on}}$) from $-$80$\,\rm km\,s^{-1}$ to $160\,\rm km\,s^{-1}$ averaged over an area of $180\arcsec\times180\arcsec$ centered in $\ell=19.9^{\circ}, b=-0.5^{\circ}$. The dashed red curve is a second-order polynomial fit ($T_{\mathrm{off}}$) to the absorption-free channels of the \ion{H}{i} spectrum at 33.5--$43.0\rm\,km\,s^{-1}$ and 56.0--$65.5\rm\,km\,s^{-1}$ (see Sect.~\ref{sec:poly_fitting}). We estimated the HISA spectrum by then subtracting the \ion{H}{i} spectra from the fitted background emission. The GRS \element[ ][13]{CO} spectrum \citep{2006ApJS..163..145J} covering velocities from $-$5$\,\rm km\,s^{-1}$ to $135\,\rm km\,s^{-1}$ is shown in blue and has been multiplied by a factor of ten for better visibility.}
        \label{fig:whole_HI_spectrum}
    \end{figure}

    The random motion of individual \ion{H}{i} clouds, superposed on the Galactic rotation, contributes significantly to the broadening of the observed 21cm emission and creates multiple emission peaks as seen in Fig.~\ref{fig:whole_HI_spectrum}. \ion{H}{i} spectra show many features and \element[ ][13]{CO} line emission has to be inspected to search for HISA. Since we assume that the CNM is associated with cold molecular gas, we take local \element[ ][13]{CO} emission peaks as a reference point. We thus identify HISA by constraining these features kinematically. The \element[ ][13]{CO} emission peaks at different velocities are not associated with GMF20.0-17.9 as their velocities are attributed to neighboring spiral arm structures \citep[e.g.,][]{2008AJ....135.1301V,2014ApJ...783..130R,2019ApJ...885..131R}.
    For the analysis of the physical properties of HISA, we followed the derivation by \citet{2000ApJ...540..851G} and \citet{2020A&A...634A.139W}. 
    A comprehensive discussion of the radiative transfer of HISA clouds is given in \citet{2000ApJ...540..851G}, \citet{2003ApJ...598.1048K}, and \citet{2003ApJ...585..823L}. Adopting the geometric model from \citet{2000ApJ...540..851G}, we identify four different cloud components when looking toward a HISA cloud, which we describe below.
    
    According to this model \citep[see Fig.~2 in][]{2020A&A...634A.139W}, we observe emitting foreground and background clouds that have spin temperatures of $T_{\mathrm{fg}}$ and $T_{\mathrm{bg}}$, respectively. Between these clouds a cold absorbing HISA cloud can be located, with a spin temperature of $T_{\mathrm{HISA}}$. Diffuse continuum emission, $T_{\mathrm{cont}}$, is assumed to be in the background. Strong continuum point sources will be neglected as they contaminate the absorption features that are caused by HISA.
    
    By comparing an ``$\mathrm{on}$'' spectrum, where a HISA cloud is located along the line of sight, with the ``$\mathrm{off}$'' spectrum that we would observe in the absence of the HISA cloud, we can derive the optical depth of the HISA component \citep[see e.g., Eq.~(6) in][]{2000ApJ...540..851G}, defined as
    
    \begin{equation}
        \tau_{\mathrm{HISA}} = -\mathrm{ln}\left(1-\frac{T_{\mathrm{on}}-T_{\mathrm{off}}}{T_{\mathrm{HISA}} - pT_{\mathrm{off}} - T_{\mathrm{cont}}}\right) \: ,
        \label{equ:T_ON-T_OFF}
    \end{equation}{}
    
    \noindent with the dimensionless parameter $p\equiv T_{\mathrm{bg}}\,\left(1-e^{-\tau_{\mathrm{bg}}}\right)/T_{\mathrm{off}}$ describing the fraction of background emission in the optically thin limit \citep{1993A&A...276..531F}. Assuming a HISA spin temperature $T_s$ ($=T_{\rm HISA}$), we can then calculate the \ion{H}{i} column density of the cold \ion{H}{i} gas using the general form \citep{2013tra..book.....W}
    
    \begin{equation}
        \frac{N_{\mathrm{H}}}{\rm cm^{-2}} = 1.8224\times 10^{18}\,\, \frac{T_s}{\rm K}\,\int\tau\left(T_s,v\right)\,\left( \frac{\mathrm{d}v}{\rm km\,s^{-1}}\right) \: ,
        \label{equ:HI_column_density}
    \end{equation}{}
    
    \noindent where $T_s$ is the spin temperature of atomic hydrogen and $\tau\left(T_s,v\right)$ describes the optical depth.
    
    {}

    To reliably identify HISA features, it is crucial to know the emission in the absence of a HISA cloud. Many methods have been tested to estimate $T_{\mathrm{off}}$ in Eq.~\eqref{equ:T_ON-T_OFF} \citep[e.g.,][]{2000ApJ...540..851G,2003ApJ...598.1048K,2003ApJ...585..823L,2008ApJ...689..276K,2020A&A...634A.139W}. \citet{2020A&A...634A.139W} tested estimating the background spectrum $T_{\mathrm{off}}$ by measuring several $\mathrm{off}$ positions offset from apparent absorption features at slightly shifted lines of sight. Their spectra partly show large variations depending on the line of sight, so the assumption that the \ion{H}{i} background emission stays spatially constant does not really hold. Therefore, we refrain from selecting actual $\mathrm{off}$ positions to estimate $T_{\mathrm{off}}$. Instead, we estimate $T_{\mathrm{off}}$ for each line of sight by fitting the baselines of absorption features with polynomial functions. The fits reconstruct an $\mathrm{off}$ spectrum as if there was no absorption present. Different studies have been conducted successfully by applying polynomial fitting procedures \citep{2003ApJ...598.1048K,2003ApJ...585..823L,2020A&A...634A.139W}.
    
    We extensively tested various methods to find an independent and systematic fitting procedure. Fitting the baselines with first and second order polynomial functions yielded the most robust results as these functions are not sensitive to small-scale fluctuations along the spectral axis. We therefore rebinned the spectral axis of \ion{H}{i} emission by a factor of two, which gave the best results for reconstructing $T_{\mathrm{off}}$, independent of the chosen velocities at which the baselines were fitted. Higher-order polynomial functions are prone to either over- or underestimating the background spectrum. As outlined below, we utilized a combination of first and second order polynomials in order to fit the baselines of HISA spectra. For the baseline fitting, we furthermore smoothed the \ion{H}{i} emission maps spatially to an angular resolution of $\Delta\Theta=80\arcsec$ to enhance the brightness sensitivity.
    
\begin{figure}[!htbp]
        \centering
            \resizebox{\hsize}{!}{\includegraphics{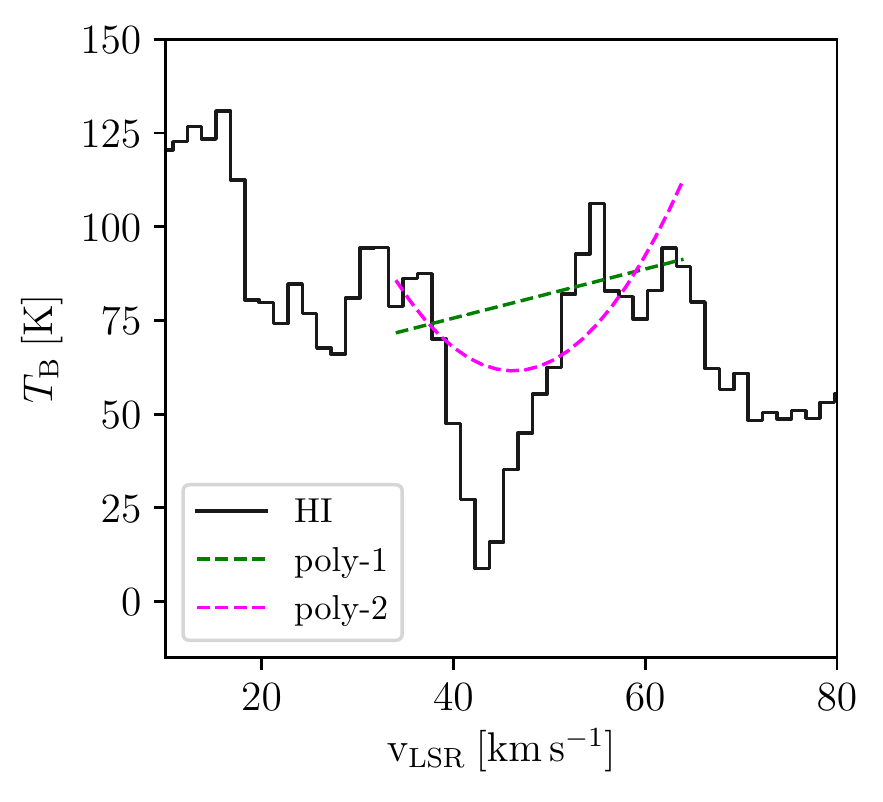}}
        \caption{Comparison of baseline fits toward continuum emission. The black curve shows an example spectrum of \ion{H}{i} emission centered in $\ell=18.95^{\circ}, b=-0.03^{\circ}$ that is contaminated by continuum emission. The dashed magenta and green curve show a second and first order polynomial fit to the velocity channels of the \ion{H}{i} spectrum at 33.5--$43.0\rm\,km\,s^{-1}$ and 56.0--$65.5\rm\,km\,s^{-1}$, respectively. Due to the continuum contamination, the second order polynomial yields a bad fit to the HISA baseline.}
        \label{fig:poly12_fit}
    \end{figure}

    Irrespective of the actual presence of \element[][13]{CO} emission at individual positions, every pixel spectrum is searched for HISA and fitted at the velocities $33.5-43.0\rm\,km\,s^{-1}$ and $56.0-65.5\rm\,km\,s^{-1}$, omitting the velocity range associated with GMF20.0-17.9. In the first cycle of the fitting procedure, all spectra are fitted with second order polynomial functions ($f(x)=ax^2+bx+c$). Spectra that are contaminated by continuum emission produce bad second order polynomial fits, with $a>0$. For those spectra, we used first order fits instead. Figure~\ref{fig:poly12_fit} presents a comparison between first and second order polynomial fits toward a position that is contaminated by diffuse continuum emission, which can contribute to the broadening of the absorption profile.
    
    {}
    
\begin{figure}[!htbp]
      \centering
        \resizebox{\hsize}{!}{\includegraphics{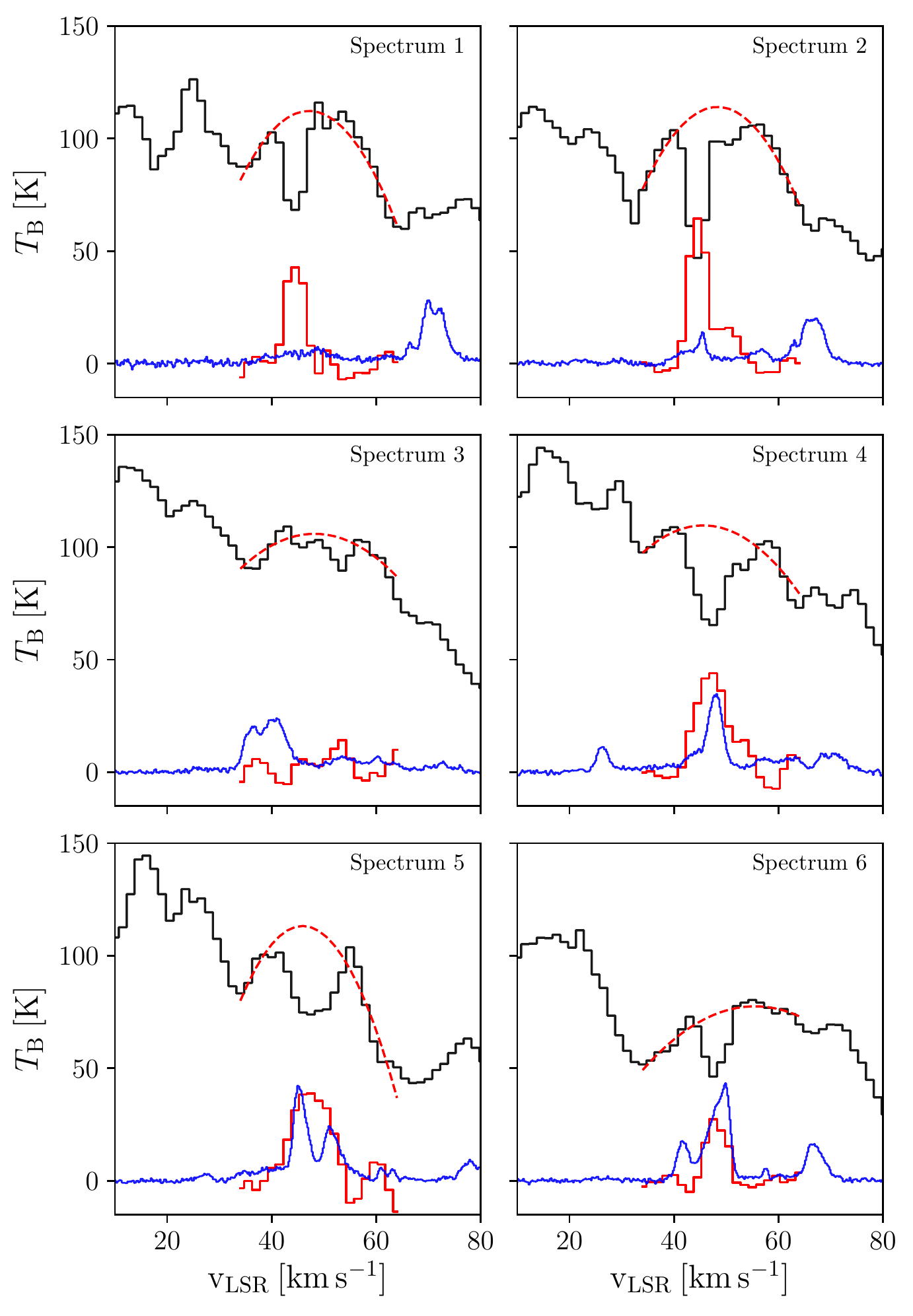}}
      \caption[]{\ion{H}{i} ($T_{\mathrm{on}}$) and extracted HISA spectra ($T_{\mathrm{off}}-T_{\mathrm{on}}$) toward the regions marked by the yellow circles in the bottom panel of Fig.~\ref{fig:HI_moment0_with_regions}. The colors are the same as described in Fig.~\ref{fig:whole_HI_spectrum}.}
      \label{fig:baseline_fits}
    \end{figure}

    Figure~\ref{fig:baseline_fits} shows our baseline fitting procedure and extracted HISA spectra from example regions of GMF20.0-17.9. The example regions have been selected based on the visual inspection of the \ion{H}{i} and \element[][13]{CO} emission maps. The spectra show the case of HISA with strong, weak, and no molecular counterparts as well as no HISA at all. The final HISA maps were inferred by subtracting the native THOR \ion{H}{i} emission with a spatial and spectral resolution of 40\arcsec and $1.5\rm\,km\,s^{-1}$, respectively, from the fitted baselines. The rms noise of the extracted HISA spectra is $\sim$8$\rm\,K$ and arises from the noise of the observations and the uncertainty of the fitting procedure. We discuss these uncertainties in Appendix~\ref{sec:discussion_extraction_method}.
    
    Using this approach, we are biased in the search of HISA since we utilize \element[ ][13]{CO} velocities to constrain the velocities of extracted HISA features. We lack a systematic approach to find HISA independent of molecular line emission.
    At the spectral resolution of $1.5\rm\,km\,s^{-1}$, we are not able to detect narrow self-absorption features \citep[HINSA;][]{2003ApJ...585..823L,2008ApJ...689..276K} that can be identified through line profile characteristics, such as the line width and the second derivative of the absorption feature. \ion{H}{i} self-absorption features with line widths $\geq 1\rm\,km\,s^{-1}$ are difficult to differentiate from emission troughs. The kinematic information of molecular line emission is therefore crucial in our analysis.
    
\section{Results}
   
\subsection{\ion{H}{i} self-absorption}
    In order to compare the kinematics in a statistical sense, we regridded the HISA data to the same pixel scale as the \element[][13]{CO} GRS data. The properties and kinematics of the CNM were analyzed by fitting single Gaussian components to the HISA spectra. Due to the limited velocity resolution, it is not feasible to resolve multiple HISA components between $43$ and $56\rm\,km\,s^{-1}$. Fits that have a peak intensity of $>25\rm\, K$ ($\sim$3$\sigma$) and a line width between $1.5$ and $20\rm\,km\,s^{-1}$ (FWHM) are considered good.
    
\begin{figure*}[!htbp]
      \centering
        \includegraphics[width=1.0\textwidth]{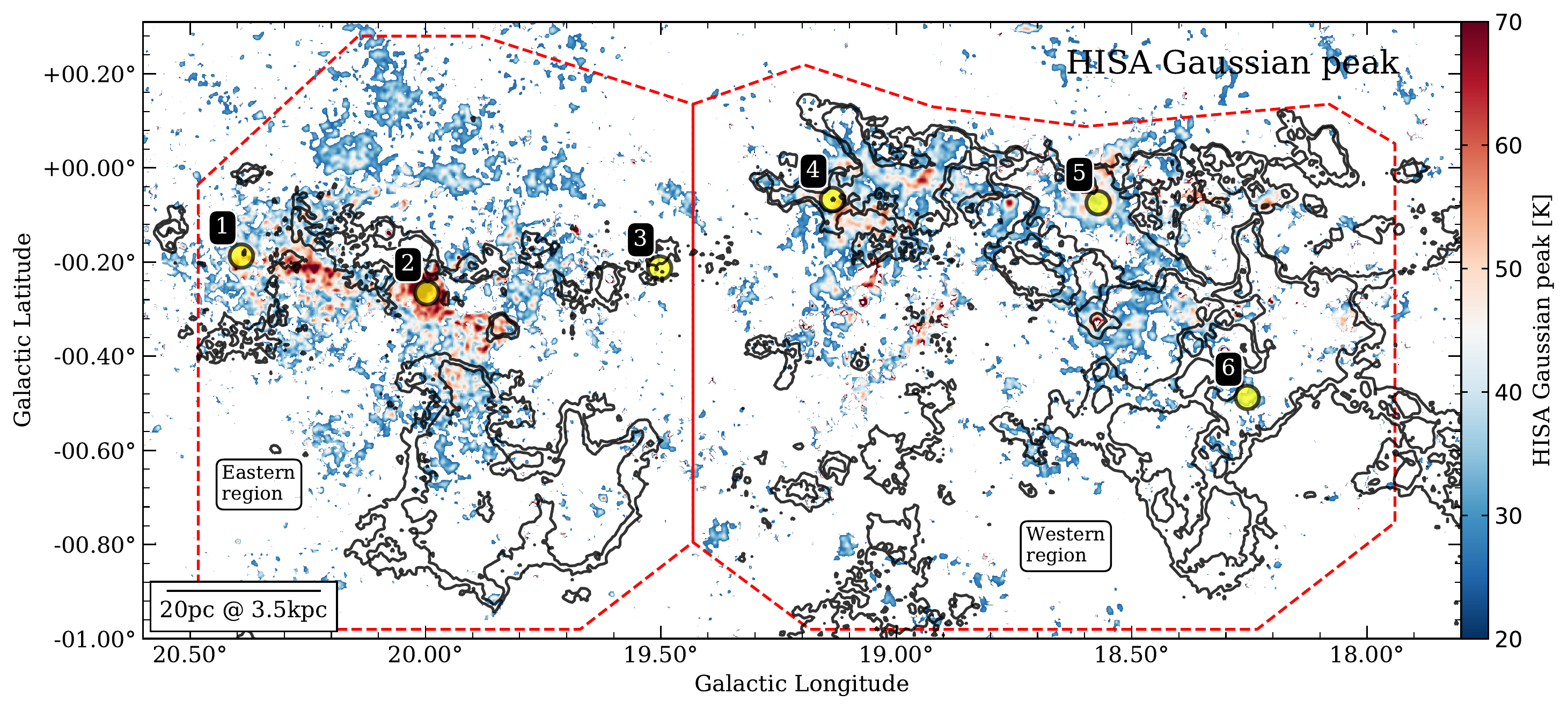}
      \caption[]{Peak values of a Gaussian fit applied to the estimated HISA spectra. This panel presents the fitted peak values of the extracted HISA spectra. The spectra have been fitted with a single-component Gaussian curve. The black contours represent the integrated \element[ ][13]{CO} emission from 42 to $57\rm\,km\,s^{-1}$ at levels of $10.5$ and $15\,\rm K\,km\,s^{-1}$. The red dashed polygons define the eastern and western part of the filament. The yellow circles mark the regions of the spectra shown in Fig.~\ref{fig:baseline_fits}.}
      \label{fig:HISA_gauss_amp}
    \end{figure*}

    The fitted peak values of the extracted HISA spectra are shown in Fig.~\ref{fig:HISA_gauss_amp}. The derived HISA peaks have intensities between $\sim$30$\rm\,K$ and $\sim$70$\rm\,K$. By comparing the inferred HISA features with the molecular gas emission, the filament can be separated into two subregions (see Fig.~\ref{fig:HISA_gauss_amp}). The western part of the filament ($19.3^{\circ}\gtrsim\ell\gtrsim 17.9^{\circ}$) shows good spatial correlation between HISA and \element[ ][13]{CO} as the cold atomic gas is expected to be closely associated with its molecular counterpart. We assess the spatial correlation quantitatively in Sect.~\ref{sec:HOG} to confirm this finding. However, the eastern part of the filament ($20.5^{\circ}\gtrsim\ell\gtrsim 19.5^{\circ}$) shows significant HISA that does not spatially overlap with the \element[ ][13]{CO} emission at the velocities around $\sim$45$\rm\,km\,s^{-1}$. On the eastern side of the cloud, the CNM as traced by HISA appears to envelop the denser molecular filament. The extracted features indicate the presence of a cold \ion{H}{i} cloud as the velocities generally agree with the molecular gas (Fig.~\ref{fig:peak_velocity_map}). Furthermore, optical depth measurements against bright continuum sources reveal high optical depths in the same velocity regime (Sect.~\ref{sec:HI_emission_optical_depth}). This underlines the robustness of the extraction method. We examined the two subregions separately in the following analysis.
    
\subsection{Kinematics}
\begin{figure*}[!htbp]
      \centering
        \includegraphics[width=1.0\textwidth]{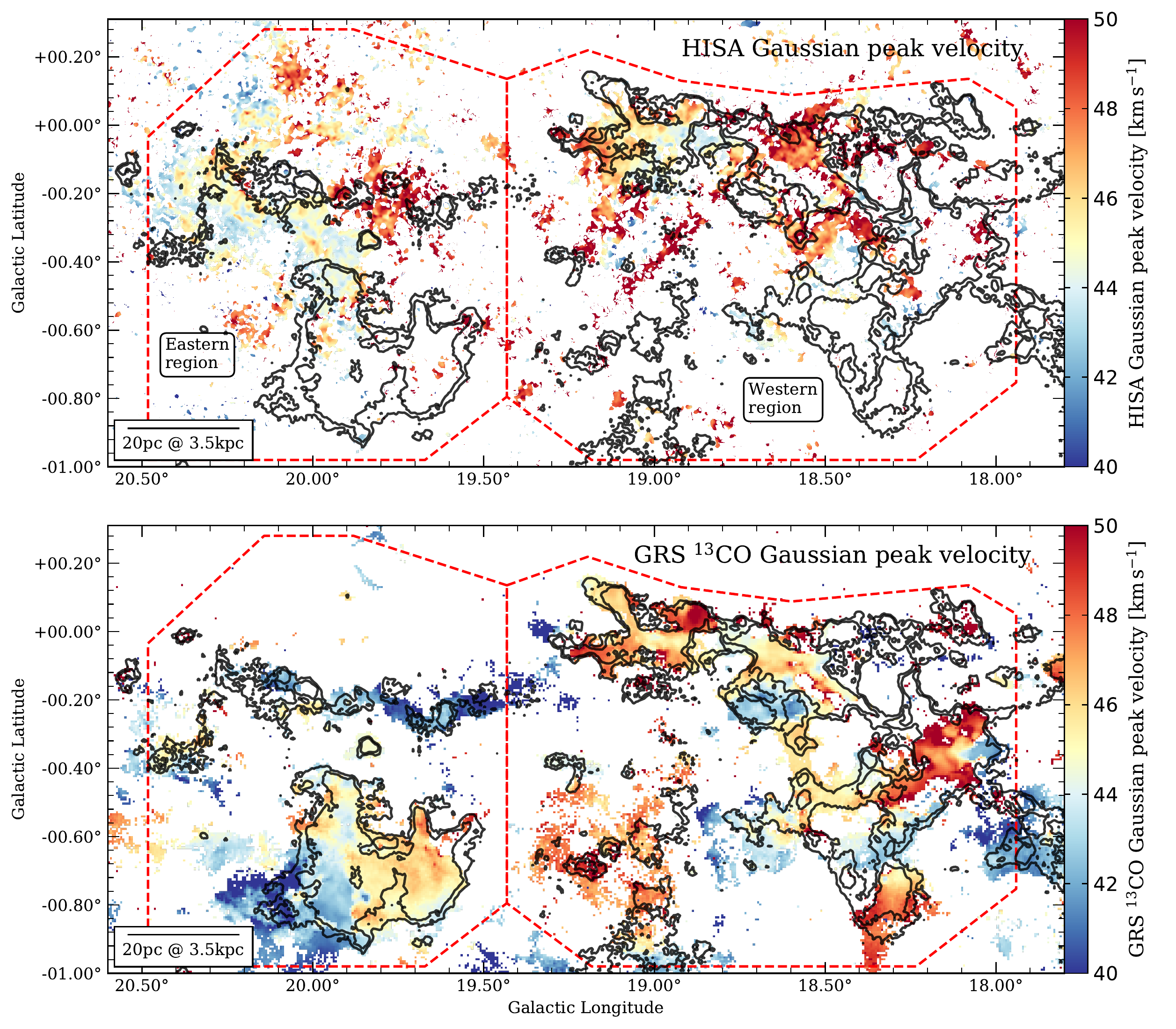}
      \caption[]{Peak velocity maps. The upper panel presents the HISA peak velocities inferred from Gaussian fits. The lower panel shows the fitted \element[ ][13]{CO} peak velocities. The black contours indicate the integrated \element[ ][13]{CO} emission at levels of $10.5$ and $15\rm\,K\,km\,s^{-1}$ for reference. The red dashed polygons mark the eastern and western part of the filament that were used for a separate analysis.}
      \label{fig:peak_velocity_map}
    \end{figure*}

     We smoothed the \element[ ][13]{CO} spectra to a spectral resolution of $1.5\rm\,km\,s^{-1}$ and applied single-component Gaussian fitting to be consistent in our analysis. Emission features with a peak intensity of $>1.25\rm\,K (\sim 5\sigma)$ and a line width $1.5\rm\,km\,s^{-1}<FWHM<20\,km\,s^{-1}$ are considered to be good fits. The peak velocity maps of HISA and \element[ ][13]{CO} are presented in Fig.~\ref{fig:peak_velocity_map}. The peak velocities of HISA in the eastern part of the filament show a velocity of $\sim$$44-46\rm\,km\,s^{-1}$. The western part reveals slightly higher peak velocities from $\sim$45 to $\sim$49$\rm\,km\,s^{-1}$. The peak velocities of \element[ ][13]{CO} show a coherent distribution along the filament \citep{2014A&A...568A..73R}.
    
\begin{figure*}[!htbp]
      \centering
        \includegraphics[width=1.0\textwidth]{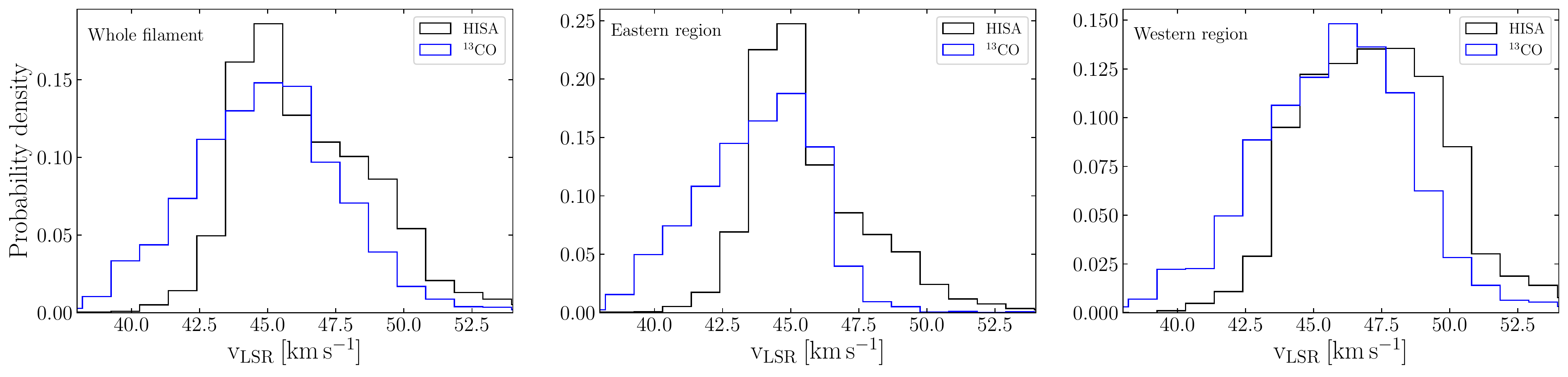}
      \caption[]{Histogram of peak velocities. The histograms show the peak velocities of HISA and \element[ ][13]{CO} in black and blue, respectively. The middle and right panels show the velocity distribution in the regions marked by the left and right polygon in Fig.~\ref{fig:peak_velocity_map}, respectively.}
      \label{fig:histogram_peak_velocity}
    \end{figure*}

\begin{figure*}[!htbp]
      \centering
        \includegraphics[width=1.0\textwidth]{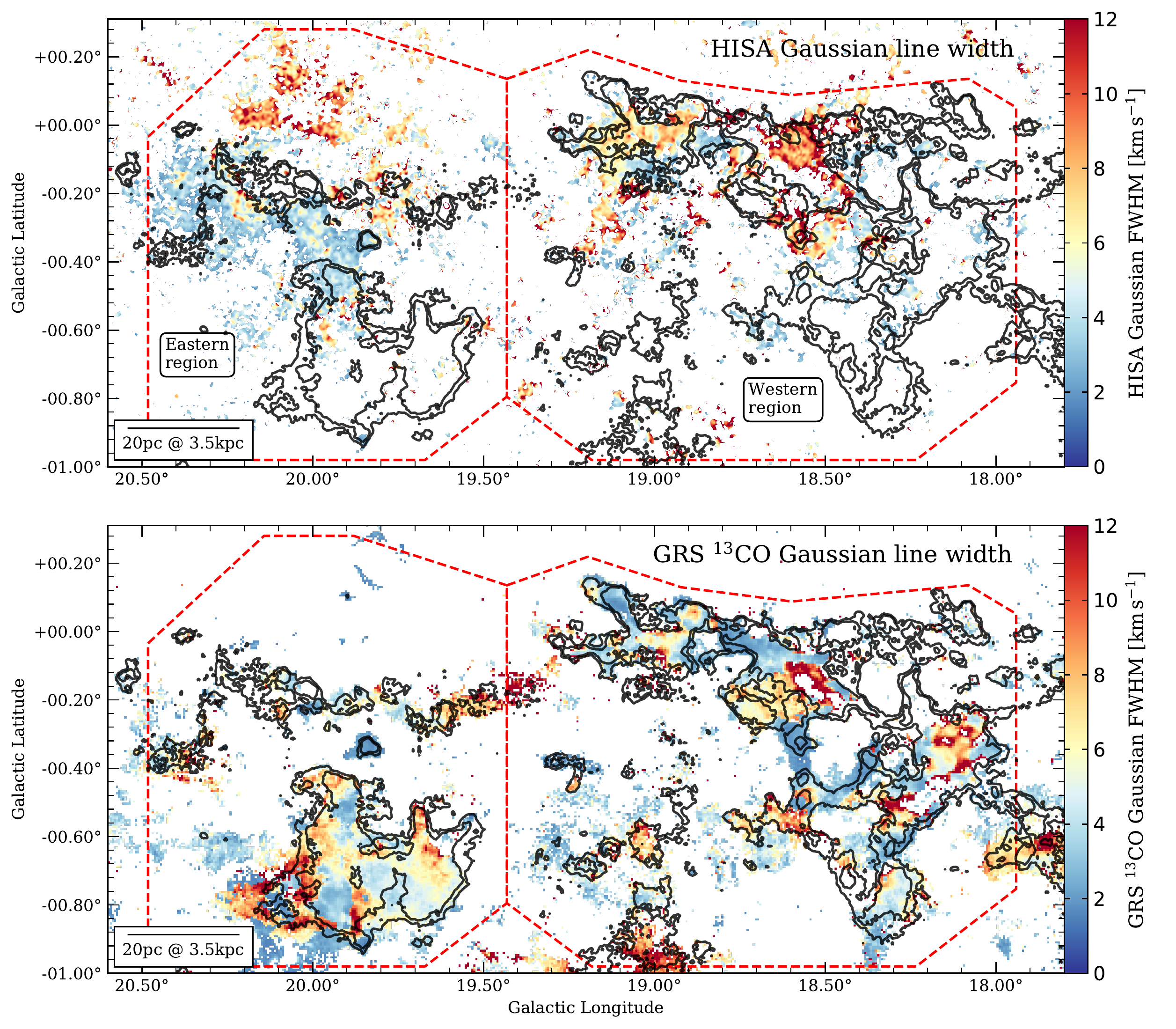}
      \caption[]{Line width maps. The upper panel presents the HISA line widths inferred from Gaussian fits. The lower panel shows the fitted \element[ ][13]{CO} line widths. The black contours indicate the integrated \element[ ][13]{CO} emission at levels of $10.5$ and $15\rm\,K\,km\,s^{-1}$ for reference. The red dashed polygons mark the eastern and western part of the filament that were used for a separate analysis.}
      \label{fig:HISA_gauss_linewidths}
    \end{figure*}

    Although there are slight systematic differences in peak velocity at some positions, the median of the histograms of peak velocities reveal good agreement between \ion{H}{i} and \element[][13]{CO} emission in both the eastern and western regions (Fig.~\ref{fig:histogram_peak_velocity}). The similar velocities are a confirmation that the extracted HISA structures are trustworthy, even though HISA and \element[ ][13]{CO} show a lower degree of line-of-sight correlation in the eastern part of the filament.
    The HISA structures in the northern part of the eastern region reveal large line widths of $\sim$8--$10\rm\,km\,s^{-1}$ (Fig.~\ref{fig:HISA_gauss_linewidths}). The bulk of HISA south of the \element[ ][13]{CO} contours shows line widths of $3$--$6\rm\,km\,s^{-1}$. Possible implications of this line width enhancement are discussed in Sect.~\ref{sec:discussion_kinematics}.
    
\begin{figure*}[!htbp]
      \centering
        \includegraphics[width=1.0\textwidth]{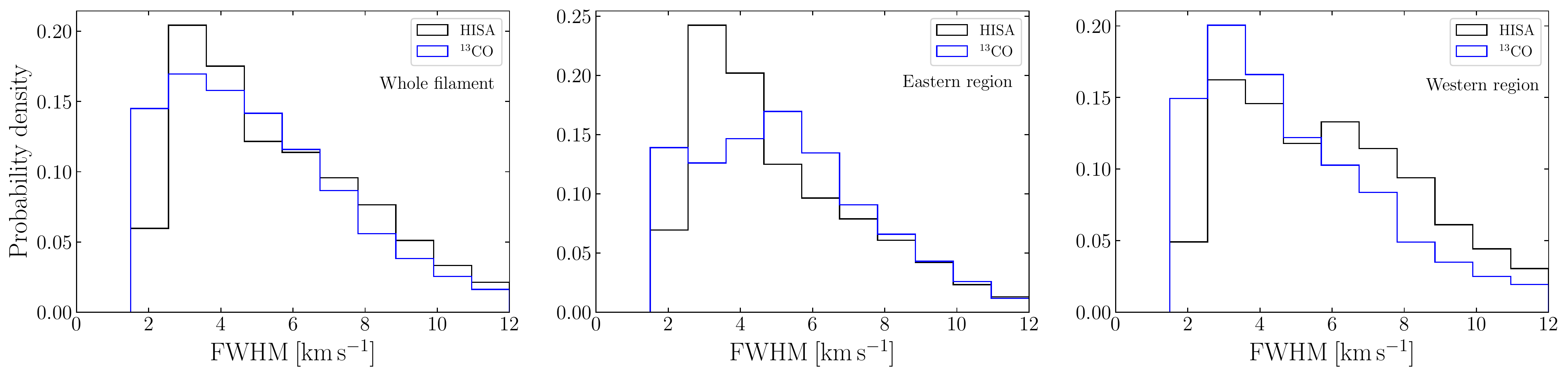}
      \caption[]{Histogram of line widths. The histograms show the line widths of HISA and \element[ ][13]{CO} in black and blue, respectively. The middle and right panels show the line width distribution in the regions marked by the left and right polygon in Fig.~\ref{fig:HISA_gauss_linewidths}, respectively.}
      \label{fig:histogram_linewidths}
    \end{figure*}

    The \element[ ][13]{CO} line widths are $\sim$2--$3\rm\,km\,s^{-1}$ in the western part and show line widths that are slightly higher in the eastern part (Fig.~\ref{fig:histogram_linewidths}).
    Assuming a kinetic temperature, we can estimate the expected thermal line width. In local thermodynamic equilibrium (LTE), the thermal line width (FWHM) is given by $\Delta v_{\mathrm{th}} = \sqrt{8\,\mathrm{ln}2\,k_BT_k/(\mu m_{\rm H})}$, where $k_B$, $T_k$, and $\mu$ are the Boltzmann constant, kinetic temperature, and the mean molecular weight of \ion{H}{i} and the CO molecule in terms of the mass of a hydrogen atom $m_{\rm H}$, respectively. If different line broadening effects are uncorrelated, the total observed line width will be
    
    \begin{equation}
        \Delta v_{\mathrm{obs}} = \sqrt{\Delta v_{\mathrm{th}}^2 + \Delta v_{\mathrm{nth}}^2 + \Delta v_{\mathrm{res}}^2} \: ,
    \end{equation}{}
    
    \noindent where $\Delta v_{\mathrm{nth}}$ is the line width due to nonthermal effects and $\Delta v_{\mathrm{res}}$ is the line width introduced by our spectral resolution and is equal to $1.5\rm\,km\,s^{-1}$.
    
    The observed \element[ ][13]{CO} line widths even at the lower end of the distribution at $2$--$3\rm\,km\,s^{-1}$ cannot be explained by thermal line broadening. Effects such as turbulent motions are most likely the dominant driver for the broadening of the \element[ ][13]{CO} line. More than 70\% of the observed HISA line widths are $\geq 3\rm\,km\,s^{-1}$.
    
    We can investigate the three-dimensional Mach number of the filament by assuming isotropic turbulence $\mathcal{M}=\sqrt{3}\,\sigma_{\mathrm{nth}}/c_s$, where $\sigma_{\mathrm{nth}}$ is the nonthermal one-dimensional velocity dispersion that is related to the nonthermal line width via $\Delta v_{\mathrm{nth}}=\sqrt{8\,\mathrm{ln}2}\,\sigma_{\mathrm{nth}}$. The sound speed $c_s$ is estimated using a mean molecular weight $\mu=2.34$ for the molecular gas and $\mu=1.27$ for the cold \ion{H}{i} phase \citep{1973asqu.book.....A,2000asqu.book.....C}. To calculate the thermal component of the velocity dispersion, we assumed the spin temperature of the cold atomic hydrogen to be close to the kinetic temperature and set $T_k=T_{\mathrm{HISA}}=40\rm\,K$. As we find \element[][13]{CO} excitation temperatures as high as $\sim$25$\rm\,K$ where the line is becoming optically thick (see Sect.~\ref{sec:CO_column_dens}), we assumed that the actual kinetic temperature of \element[][13]{CO} is close to the excitation temperature in those regions, meaning these regions are dense and in LTE. We therefore set a uniform kinetic temperature of $T_k=20\rm\,K$ for the Mach number estimates of \element[][13]{CO}.
    
    \begin{figure}[!htbp]
      \centering
        \resizebox{\hsize}{!}{\includegraphics{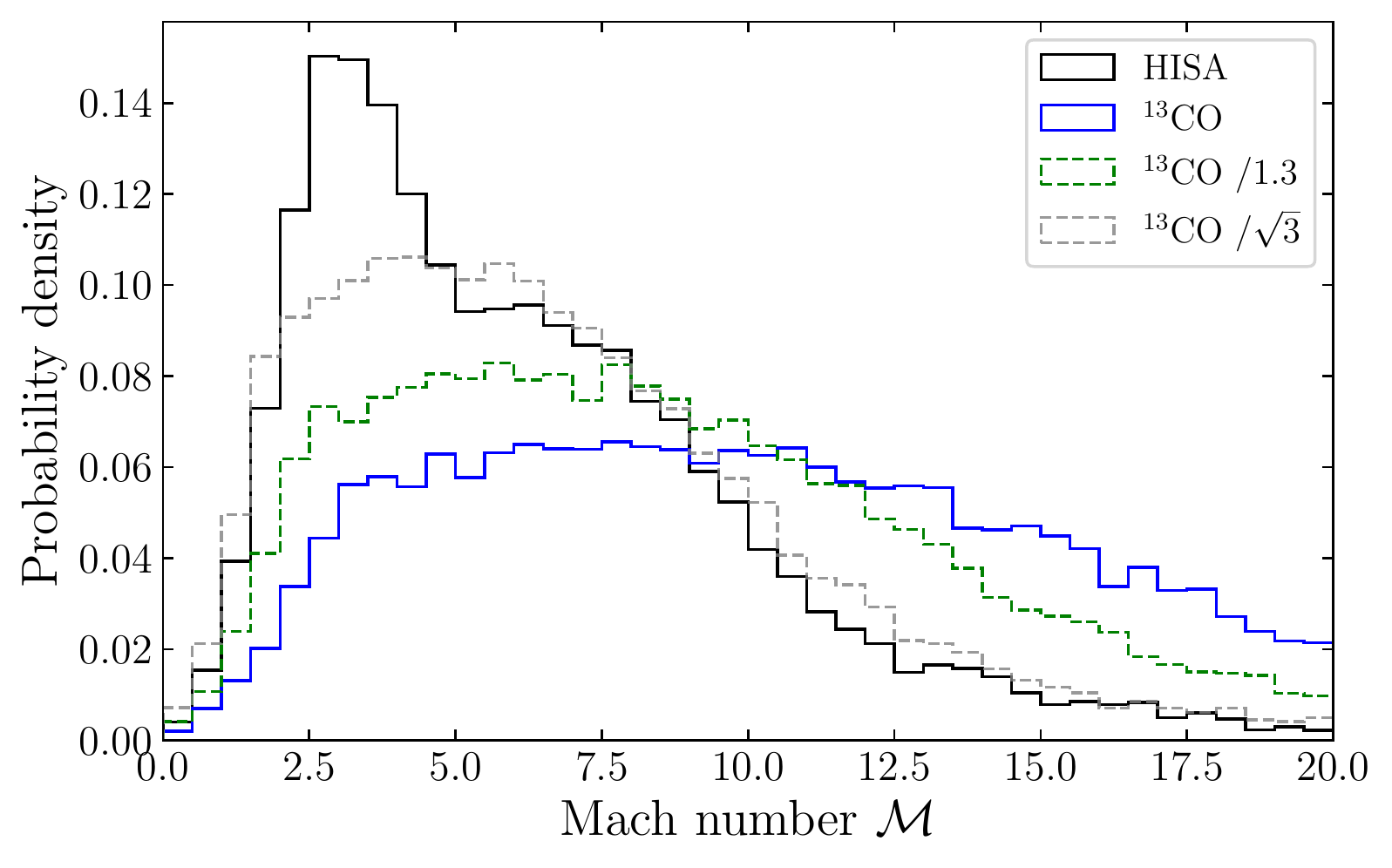}}
      \caption[]{Distribution of the turbulent Mach number across the whole filament. The cold atomic hydrogen traced by HISA is shown in black. The Mach numbers of the molecular gas seen in \element[][13]{CO} emission are shown in blue. The green and gray dashed distribution shows the \element[][13]{CO} Mach numbers if the \element[][13]{CO} line width is on average overestimated by 30\% and $\sqrt{3}$, respectively.}
      \label{fig:mach_number}
    \end{figure}
    Figure~\ref{fig:mach_number} shows that the Mach number of the CNM traced by HISA peaks at $\sim$3, indicating that a significant fraction of the CNM has transonic and supersonic velocities. Furthermore, there is an indication of a shoulder at $\mathcal{M}\sim6$. The Mach numbers of \element[][13]{CO} show a broad distribution and are dominated by supersonic motions. The distribution is slightly skewed toward higher Mach numbers as we observe multiple \element[][13]{CO} components between 43 and $56\rm\,km\,s^{-1}$ in some regions. Consequently, fitting single Gaussian components results in overly broad line widths where we observe multiple velocity components. Hence, the nonthermal velocity dispersion and therefore the Mach number is systematically overestimated. If we utilize a single Gaussian component fit on the GRS spectra at the full spectral resolution of $0.21\rm\,km\,s^{-1}$, the \element[][13]{CO} Mach number distribution does not significantly change. Therefore, the spectral smoothing has a negligible effect on the derivation of the Mach numbers if we fit single components. To address the uncertainty of possible component blending, we assumed that the \element[][13]{CO} line width is on average systematically overestimated by 30\% due to the single-component fitting. The Mach number distribution is then shifted to lower values with a more pronounced peak (see Fig.~\ref{fig:mach_number}). Furthermore, if the filament lacks spatial isotropy, we could be overestimating the Mach number by a factor as high as $\sqrt{3}$, which would lead to a distribution with a median of $\mathcal{M}\sim6$ (Fig.~\ref{fig:mach_number}).
\subsection{Column density and mass}
\subsubsection{CNM column density traced by HISA}
    We calculated the column density of the CNM traced by HISA following Eqs.~\eqref{equ:T_ON-T_OFF} and \eqref{equ:HI_column_density}. We therefore have to assume either an optical depth or a spin temperature. As we know that HISA traces the coldest component of atomic hydrogen, we assumed a constant spin temperature of $T_S=40\rm\,K$ for the whole cloud to calculate the column density. This is a typical spin temperature of cold self-absorbing \ion{H}{i} clouds \citep[e.g.,][]{1974AJ.....79..527K,2000ApJ...540..851G,2003ApJ...586.1067H}. We emphasize that a constant spin temperature is an approximation that might not hold for every region of the cloud. However, the maximum spin temperature is constrained in Appendix~\ref{sec:discussion_max_spin_temperature} and the actual temperature variation should be moderate. Different spin temperatures (if constant over the whole cloud) will not change the structure of the column density distribution in the cloud but only change the normalization factor.
    
    Furthermore, we have to assume the fraction of background emission parameterized by the factor $p$ (Eq.~\ref{equ:T_ON-T_OFF}). Although we cannot measure this parameter directly, we can constrain $p$ by its effect on the spin temperature and the location of the cloud. Because of the cloud's location toward the inner Galactic plane ($\ell\sim 19^{\circ}$) and its distance of $\sim$3.5$\rm\,kpc$ \citep{2014A&A...568A..73R}, it is unlikely that most of the \ion{H}{i} emission originates in the foreground. The fraction of background emission should therefore be at least $p\gtrsim 0.5$. Since self-absorption can also be induced by \ion{H}{i} emission from the far side of the Galaxy due to the kinematic distance ambiguity, the background fraction $p$ should be systematically higher than the foreground emission fraction. Therefore, we assumed a background fraction of $p=0.9$ and discuss its uncertainties in Appendix~\ref{sec:discussion_background_fraction}. \citet{2020A&A...634A.139W} assumed the same background fraction for their HISA analysis of the giant molecular filament GMF38.1-32.4a \citep{2014A&A...568A..73R}. Furthermore, \citet{2000ApJ...540..851G} argue that the HISA detection is biased toward higher $p$ values since a high background fraction is more efficient in producing prominent HISA features.
    
\subsubsection{Molecular gas column density traced by \element[ ][13]{CO}}\label{sec:CO_column_dens}
    In the optically thin limit, the \element[ ][13]{CO} column density is computed by \citep{2013tra..book.....W}
    
    \begin{equation}
     N(\element[ ][13]{CO}) = 3.0\times 10^{14}\,\frac{\int T_B(v)\,\mathrm{d}v}{1-\mathrm{exp}(-5.3/T_{\mathrm{ex}})} \: ,
     \label{equ:N_CO}
    \end{equation}{}
    
    \noindent where $N(\element[ ][13]{CO})$ is the column density of \element[][13]{CO} molecules in $\rm cm^{-2}$, $\mathrm{d}v$ is in units of $\rm km\,s^{-1}$, $T_B$ and $T_{\mathrm{ex}}$ are the brightness temperature and excitation temperature of the \element[ ][13]{CO} line in units of Kelvin, respectively. By assuming that the excitation temperatures of \element[ ][12]{CO} and \element[ ][13]{CO} are the same in LTE, we derived the excitation temperature from \element[ ][12]{CO}(1-0) emission data of the FOREST Unbiased Galactic plane Imaging survey with the Nobeyama 45m telescope \citep[FUGIN;][]{2017PASJ...69...78U}, using \citep{2013tra..book.....W}
    \begin{equation}
    T_{\mathrm{ex}} = 5.5\cdot\left[\textrm{ln}\left(1+\frac{5.5}{T_B^{12}+0.82}\right)\right]^{-1} \: ,
    \end{equation}{}
    \noindent where $T_B^{12}$ is the brightness temperature of the \element[][12]{CO} line in units of Kelvin. The FUGIN \element[][12]{CO} data have an angular and spectral resolution of $\Delta\Theta=20\arcsec$ and $\Delta v=1.3\rm\,km\,s^{-1}$, respectively. To calculate the excitation temperature, we reprojected the data cube on the same spatial and spectral grid as the GRS \element[][13]{CO} data of GMF20.0-17.9.
    
    We find a lower limit to the excitation temperature of $5\rm\,K$ for regions where the \element[ ][12]{CO} brightness temperature reaches the $5\sigma$ level ($2\rm\,K$). We can then derive the optical depth of the \element[][13]{CO} line from the excitation and brightness temperature, using \citep[see e.g.,][]{2013tra..book.....W,2016A&A...587A..74S}
    
    \begin{equation}
    \tau = -\mathrm{ln}\left[  1- \frac{T_B}{5.3}\cdot\left(\left[\mathrm{exp}\left(\frac{5.3}{T_{\mathrm{ex}}}\right)-1\right]^{-1}-0.16\right)^{-1}\right] \: .
    \end{equation}
    
    \noindent We estimate a lower limit of the optical depth of $\tau\sim 0.06$ for \element[ ][13]{CO} brightness temperatures above $1.25\rm\,K$ ($\sim$5$\sigma$) and the highest excitation temperatures we find ($\sim$25$\rm\,K$). Hence, we set the optical depth to $\tau=0.06$ in regions where $\tau<0.06$. Only few positions show optical depths as high as $\tau\sim 2$. We employ a correction factor to compensate for high optical depth effects by replacing the integral in Eq.~\eqref{equ:N_CO} with \citep{1982ApJ...262..590F,1999ApJ...517..209G}
    
    \begin{equation}
    \int T_B(v)\,\mathrm{d}v \rightarrow \frac{\tau}{1-e^{-\tau}}\,\int T_B(v)\,\mathrm{d}v \: .
    \end{equation}
    
    \noindent This correction factor is accurate to 15\% for $\tau<2$.
    
    To translate the \element[ ][13]{CO} column density into a column density of molecular hydrogen, we first estimated the relative abundance of \element[ ][12]{CO} with respect to \element[ ][13]{CO}. \citet{2005ApJ...634.1126M} and \citet{2014A&A...570A..65G} derived relative abundance relations based on different CO isotopologs and metallicities. At the Galactocentric radius of $D_{GC}=5.0\,\rm kpc$ \citep{2014A&A...568A..73R}, these relations give [\element[ ][12]{CO}]/[\element[ ][13]{CO}] abundances between 40 and 56. Given the large uncertainty of these numbers, we chose a canonical conversion factor of 45. The relative abundance of the main isotopolog \element[ ][12]{CO} compared to molecular hydrogen is given in \citet{2012MNRAS.423.2342F} who derive an $\rm H_2$ abundance with respect to \element[ ][12]{CO} of $X_{\element[ ][12]{CO}}^{-1}=7500$. Therefore, we adopted a conversion factor of $[\rm H_2]/[\element[ ][13]{CO}]=3.4\times 10^5$. The derived $\rm H_2$ column densities have uncertainties of at least a factor of two due to the large uncertainties in these relations. Furthermore, CO might not always be a good tracer of $\rm H_2$ as "CO-dark $\rm H_2$" could account for a significant fraction of the total $\rm H_2$ \citep{2008ApJ...679..481P,2009ApJ...692...91G,2013A&A...554A.103P,2014MNRAS.441.1628S}.
    
\begin{figure*}[!htbp]
      \centering
        \includegraphics[width=1.0\textwidth]{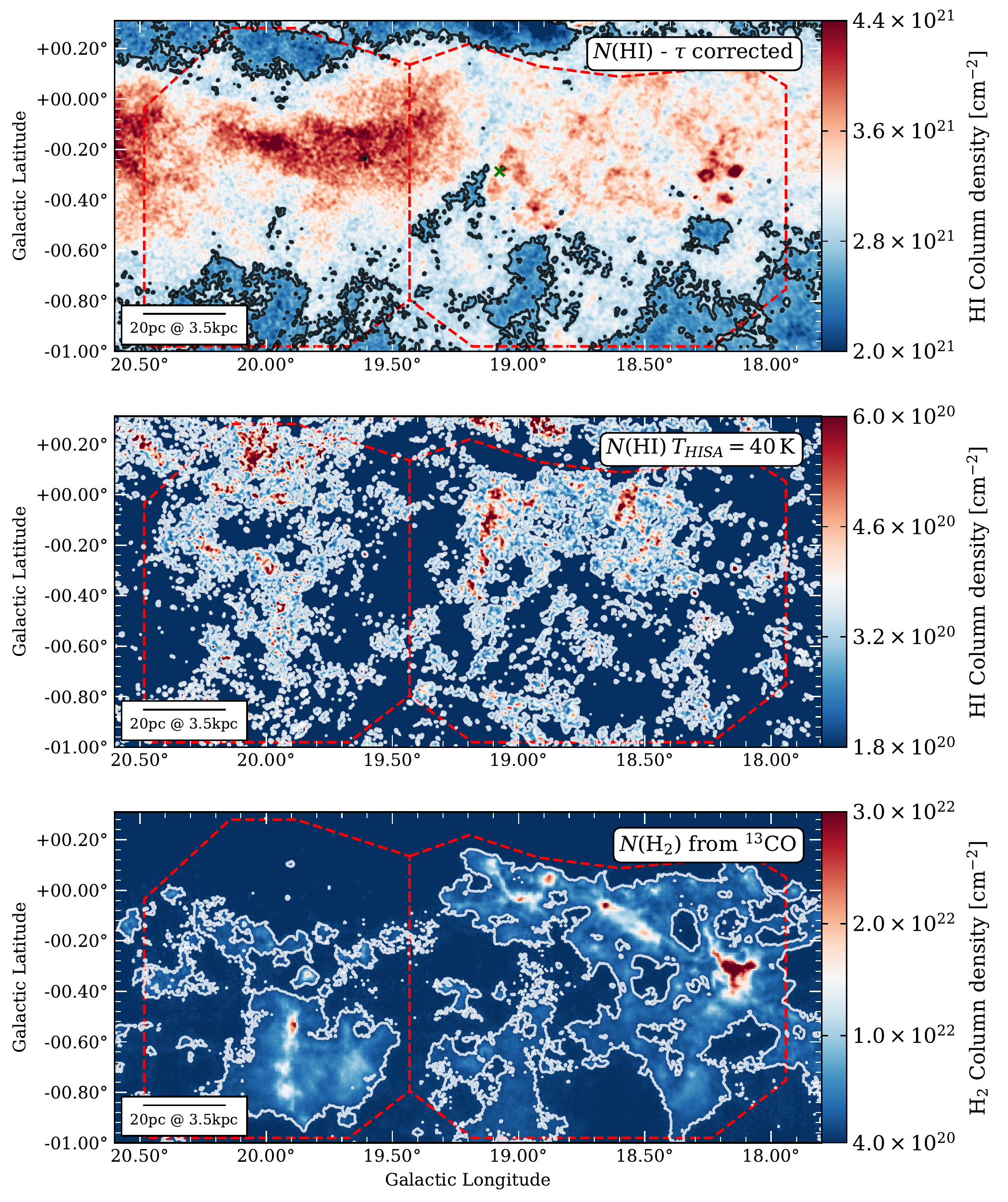}
      \caption[]{\textit{Top panel:} \ion{H}{i} column densities of the combined WNM and CNM seen in \ion{H}{i} emission. The column densities were corrected for optical depth, weak diffuse continuum emission, and the kinematic distance ambiguity. The optical depth is measured toward the continuum source G19.075-0.287 and is applied to correct the column densities throughout the whole cloud (see Sect.~\ref{sec:HI_emission_optical_depth}). The position of G19.075-0.287 is indicated by the green cross. \textit{Middle panel:} The \ion{H}{i} column densities of the CNM inferred from HISA features assuming a spin temperature of $T_{\mathrm{HISA}}=40\rm\,K$ and a background fraction of $p=0.9$. \textit{Bottom panel:} $\rm H_2$ column densities inferred from \element[ ][13]{CO} as a tracer. We assumed an $\rm H_2/\element[][13]{CO}$ ratio of $3.4\times 10^5$. The red dashed polygons in each panel indicate the eastern and western part of the filament that are analyzed separately. The white and black contours indicate the column density thresholds that were used for the derivation of the column density probability density functions (see Sect.~\ref{sec:N-PDFs}).}
      \label{fig:column_density_map}
    \end{figure*}

    For the derivation of the column densities we integrated over the whole velocity range between $43$ and $56\rm\,km\,s^{-1}$ where we find HISA. The \ion{H}{i} (CNM) and $\rm H_2$ column densities derived from HISA, and \element[ ][13]{CO}, respectively, are presented in the lower panels of Fig.~\ref{fig:column_density_map}. The column densities are partly correlated in the western part of the filament but the strongest cold \ion{H}{i} column density peaks in the eastern part ($\ell\sim 20^{\circ}$, $b\sim +0.2^{\circ}$) do not show an $\rm H_2$ column density counterpart. The strongest $\rm H_2$ column density peak in the western part ($\ell\sim 18.1^{\circ}$, $b\sim -0.3^{\circ}$) reveals little \ion{H}{i} column density but coincides with continuum emission. Continuum emission contaminates self-absorption features and hence makes it difficult to measure HISA. Most locations that are associated with continuum emission do not show HISA counterparts. However, we can measure the optical depth toward strong continuum emission sources and thus constrain the spin temperature of the HISA cloud. This is addressed in the following subsection.
    
\subsubsection{Atomic gas column density seen in \ion{H}{i} emission}\label{sec:HI_emission_optical_depth}
    In addition to HISA, we investigated the properties of atomic hydrogen (WNM+CNM) by measuring the column density from \ion{H}{i} emission and correcting for optical depth effects and diffuse continuum. We can utilize the measurement of the optical depth to constrain the spin temperature of the cold atomic hydrogen (see Appendix~\ref{sec:discussion_max_spin_temperature}). Further details of optical depth and column density corrections are given in \citet{2015A&A...580A.112B} and \citet{2020A&A...634A.139W}.
    
    As the optically thin assumption might not hold for some regions, we can utilize strong continuum emission sources to directly measure the optical depth. \ion{H}{i} continuum absorption (HICA) is a classical method to derive the properties of the CNM \citep[e.g.,][]{2004ApJ...603..560S,2003ApJ...586.1067H}. This method employs strong continuum sources, such as Galactic \ion{H}{ii} regions or active galactic nuclei (AGNs), to measure the optical depth of \ion{H}{i}. As these sources have brightness temperatures that are larger than typical spin temperatures of cold \ion{H}{i} clouds ($T_s\sim 100\rm\,K$), we observe the \ion{H}{i} cloud in absorption. The absorption feature is furthermore dominated by the CNM since the absorption is proportional to $T_s^{-1}$.
    By measuring $\mathrm{on}$ and $\mathrm{off}$ positions, we can directly compute the optical depth of \ion{H}{i}. The optical depth is given by \citep[see][]{2015A&A...580A.112B,2020A&A...634A.139W}
    
    \begin{equation}
     \tau = -\mathrm{ln}\left(\frac{T_{\mathrm{on}}-T_{\mathrm{off}}}{T_{\mathrm{cont}}}\right) \: ,
     \label{equ:HICA_tau}
    \end{equation}{}
    
    \noindent where $T_{\mathrm{on}}$ and $T_{\mathrm{off}}$ is the \ion{H}{i} brightness temperature toward a strong continuum background source and offset from the source, respectively. The brightness temperature $T_{\mathrm{cont}}$ describes the continuum level of the background source that is not affected by \ion{H}{i} absorption. The advantage of this method is the direct measurement of the optical depth. However, the HICA method requires strong continuum emission sources. As most strong continuum sources are discrete point sources, this method results in an incomplete census of optical depth measurements \citep[see][for a compilation of all optical depth measurements in the THOR survey]{2020A&A...634A..83W}. Consequently, the intrinsic structure of individual \ion{H}{i} clouds cannot be determined. Some continuum emission sources also show extended structures. Therefore, finding reliable $\mathrm{off}$ positions could pose difficulties. As we exploited THOR-only (VLA C-configuration) data for this measurement, we filter out most large-scale \ion{H}{i} emission. The THOR-only data reveal \ion{H}{i} emission of less than $30\rm\,K$, often just within the noise. Therefore, we can neglect the emission of the \ion{H}{i} cloud in Eq.~\eqref{equ:HICA_tau} and set $T_{\mathrm{off}}=0$. We can then calculate the optical depth without measuring an $\mathrm{off}$ position by
    
    \begin{equation}
     \tau_{\mathrm{simplified}} = -\mathrm{ln}\left(\frac{T_{\mathrm{on}}}{T_{\mathrm{cont}}}\right) \: .
     \label{equ:HICA_tau_simplified}
    \end{equation}
    
    \noindent Depending on the brightness of the continuum source and the \ion{H}{i} optical depth, the absorption spectrum can approach zero. Due to the noise in the spectra, the spectra can exhibit brightness temperatures smaller than zero, which is not physically meaningful. We therefore report a lower limit for the optical depth where the absorption $T_{\mathrm{on}}$ becomes smaller than $5\sigma$.
    
    Besides strong continuum sources we observe weak continuum emission throughout the Galactic plane. This component has brightness temperatures between $10$ and $50\rm\,K$.  For the derivation of the \ion{H}{i} column densities we employed the combined THOR data as in the case of HISA. The continuum emission has been subtracted during data reduction as described in Sect.~\ref{sec:methods_and_observation}. However, even weak continuum emission can still influence the observed brightness temperature. If we neglect weak continuum emission, the measured \ion{H}{i} emission will be underestimated as weak continuum emission can suppress a significant fraction of \ion{H}{i} emission \citep[e.g.,][]{2015A&A...580A.112B}. Consequently, the derived \ion{H}{i} column densities will be underestimated. We took this effect into account when computing the \ion{H}{i} column density \citep[see][Eq.~9]{2015A&A...580A.112B}. 
    
\begin{figure}[!htbp]
      \centering
        \resizebox{\hsize}{!}{\includegraphics{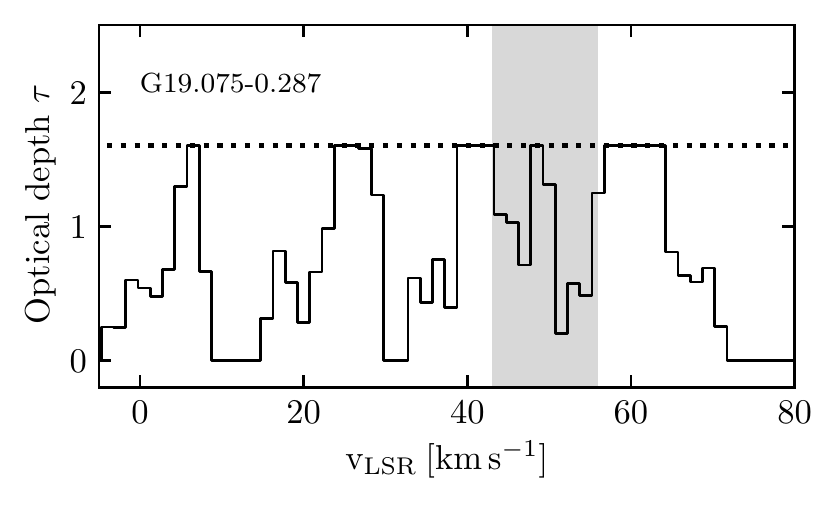}}
      \caption[]{Optical depth measurement toward the \ion{H}{ii} region \mbox{G19.075-0.287} \citep{2018A&A...619A.124W}. The plot shows the optical depth as a function of $\mathrm{LSR}$ velocity and was computed using Eq.~\eqref{equ:HICA_tau_simplified}.  For some channels, the absorption spectrum saturates and the measured optical depth is a lower limit of $\tau=1.6$, which is indicated by the horizontal dotted line. The gray shaded area indicates the velocity range between 43 and $56\rm\,km\,s^{-1}$, where HISA features have been extracted.}
      \label{fig:T_spin_tau}
    \end{figure}

    In contrast to \citet{2020A&A...634A..83W}, where they used a $6\sigma$ threshold to select continuum sources, we measured the optical depth of atomic hydrogen toward the brightest continuum sources with brightness temperatures $T_{\rm cont}>200\rm\,K$ to not suffer from low saturation limits since we expect the optical depth to be high. Four sources have been identified with this threshold. The measured optical depths of these sources vary between $0.5$ and $2.5$ (lower limit). We selected the continuum source G19.075-0.287 \citep{2018A&A...619A.124W} as a representative source for the optical depth as it is not in the $5\sigma$ saturation at most velocities between 43 and $56\rm\,km\,s^{-1}$, which gives a mean optical depth of $\tau\sim 0.9$ (Fig.~\ref{fig:T_spin_tau}). This is a reasonable approximation as the optical depth map derived by \citet{2020A&A...634A..83W} gives a mean optical depth of $\sim$1.0 when averaged over the whole filament.
    
    However, this optical depth measurement is a lower limit as G19.075-0.287 is a Galactic \ion{H}{ii} region located in the Galactic plane. As no strong extragalactic continuum sources are identified toward the position of GMF20.0-17.9, the optical depth measurement has drawbacks in the current investigation. For the emission data, we have to take into account an opacity contribution from the far side \ion{H}{i} beyond the location of the \ion{H}{ii} region. To first approximation, we assumed that the optical depth from the background is similar to that of the measured foreground. We therefore adopted $2\times\tau(v_{\rm LSR})$ for the whole map and corrected the \ion{H}{i} column density for the optical depth per velocity channel. Given the corrected mean optical depth of $2\times\tau\sim1.8$, the opacity correction factor $\tau/(1-e^{-\tau})$ increases the mean column density by a factor of $\sim$2.
    
    The derived column densities are a result of the \ion{H}{i} emission stemming from both the kinematic far ($12.0\rm\,kpc$) and near ($3.5\rm\,kpc$) side of the Milky Way. The kinematic distances have been obtained using the Kinematic Distance Utilities\footnote{\url{https://github.com/tvwenger/kd}} \citep{2018ApJ...856...52W} and employing the Galactic rotation model from \citet{2019ApJ...885..131R}. Since the distribution of the atomic gas in the Galactic plane is approximately axisymmetric with respect to the Galactic center \citep{2008A&A...487..951K}, we can assume that the atomic gas density distribution in the vertical direction is similar for a given Galactocentric radius. Using the average vertical density profile from \citet{1990ARA&A..28..215D}, we can estimate the gas fraction at the kinematic near and far side for each line of sight. Since most \ion{H}{i} emission is observed close to the Galactic midplane, the foreground gas fraction is $\sim$50\%. Therefore, due to the kinematic distance ambiguity half of the \ion{H}{i} emission is attributed to the background, which is not associated with GMF20.0-17.9. We derived the \ion{H}{i} column density map shown in the top panel of Fig.~\ref{fig:column_density_map} taking into account only the near side gas at $3.5\rm\,kpc$.
    
\subsubsection{Masses}
    As we have determined the column densities and know the distance of GMF20.0-17.9 \citep[$\sim$3.5$\rm\,kpc$,][]{2014A&A...568A..73R}, we can directly estimate the atomic and molecular mass of each part of the filament (see Table \ref{tab:masses}). All masses were calculated from the column densities integrated over 43\,--\,$56\rm\,km\,s^{-1}$.
    
    The molecular hydrogen mass of the whole filament as marked by both the red polygons in Fig.~\ref{fig:column_density_map} is 3.5$\times 10^5\rm\,M_{\odot}$. Inside the polygon regions, the mass of the total atomic hydrogen, accounting for WNM and CNM measured from \ion{H}{i} emission, corresponds to $\sim$75\% of the $\rm H_2$ mass ($2.6\times 10^5\rm\,M_{\odot}$) for the whole filament after correcting for optical depth effects, weak continuum emission, and the kinematic distance ambiguity. However, if we take into account all diffuse \ion{H}{i} emission beyond the polygon regions, arising from the region between $20.6>\ell>17.6^{\circ}$ and $-1.25<b<+0.5^{\circ}$, the mass of the total \ion{H}{i} component rises by 75\% to $\sim$4.6$\times 10^5\rm\,M_{\odot}$. The molecular filament is therefore embedded in a large gas reservoir of atomic hydrogen.
    
    The CNM mass traced by HISA corresponds to 1-5\% of the molecular mass, depending on the region and assumed spin temperature. The uncertainty of the column density directly translates to an uncertainty of mass. If we assume a spin temperature of $20\rm\,K$, instead of our canonical value of $40\rm\,K$, the mass traced by HISA decreases by a factor of $\sim$3. Hence, the largest uncertainty arises from the assumption of a spin temperature. We are able to constrain an upper limit of the spin temperature for the column density derivation by assuming an optically thick ($\tau\rightarrow\infty$) cloud, as we show in Appendix~\ref{sec:discussion_max_spin_temperature}.
    
    The atomic mass fraction generally increases toward the eastern part of the filament, agreeing with our findings in the column density distributions (see Sect.~\ref{sec:N-PDFs} for details).
    \begin{table*}[!htbp]
     \caption{Derived masses of the giant molecular filament GMF20.0-17.9.}
     \renewcommand*{\arraystretch}{1.3}
     \centering
     \begin{tabular}{ c c c c c c c c }
     \hline\hline
     Region & $M$($\rm H_2$) & $M$(\ion{H}{i}) & $M$(HISA) & $M$(HISA) & $f_{\rm HISA}$ & $f_{\rm HISA}$ &  $f_{\ion{H}{i}}$ \\
     &  &  & ($T_s=20\rm\,K$) & ($T_s=40\rm\,K$)  & ($T_s=20\rm\,K$) & ($T_s=40\rm\,K$) & \\
     & [$\rm M_{\odot}$] & [$\rm M_{\odot}$] & [$\rm M_{\odot}$] & [$\rm M_{\odot}$] & & & \\\hline
     Total & $3.5\times 10^{5}$ & $2.6\times 10^{5}$ ($4.6\times 10^{5}$)\tablefootmark{(a)} & $4.6\times 10^{3}$ & $1.3\times 10^{4}$ & 1\% & 4\% & 75\% (130\%)\tablefootmark{(a)} \\
     East & $1.1\times 10^{5}$ & $1.2\times 10^{5}$ & $1.9\times 10^{3}$ & $5.7\times 10^{3}$ & 2\% & 5\% & 110\% \\
     West & $2.3\times 10^{5}$ & $1.5\times 10^{5}$ & $2.6\times 10^{3}$ & $7.5\times 10^{3}$ & 1\% & 3\% & 65\% \\\hline
     \end{tabular}
     \tablefoot{
     The masses were calculated for each part of the filament as well as the whole filament marked by the red polygons in Fig.~\ref{fig:column_density_map}. The second column gives the molecular hydrogen mass as traced by \element[ ][13]{CO} emission. The third column shows the total atomic hydrogen mass inferred from the optical depth and continuum corrected \ion{H}{i} emission. The fourth and fifth column present the mass of the cold atomic hydrogen traced by HISA with an assumed spin temperature of 20 and $40\rm\,K$, respectively. The last three columns give the corresponding mass fractions with respect to the $\rm H_2$ mass.\\
     \tablefoottext{a}{This mass was calculated using the corrected \ion{H}{i} emission between 43 and $56\rm\,km\,s^{-1}$, $20.6>\ell>17.6^{\circ}$ and $-1.25<b<+0.5^{\circ}$.}
     }
     \label{tab:masses}
    \end{table*}
    As discussed in Appendix~\ref{sec:discussion_extraction_method}, we estimated a column density uncertainty of $\sim$40\% to account for systematic differences and noise in our baseline extraction method. Depending on the background fraction $p$, the column density further varies by a factor of $\sim$2 between $p=0.9-0.7$ (see Appendix~\ref{sec:discussion_background_fraction}).
    It is difficult to exactly quantify the uncertainty of the cold \ion{H}{i} column density and mass traced by HISA. Considering the estimated uncertainties due to the extraction method, background fraction, and spin temperature, the HISA-traced column density and mass have an uncertainty of a factor of 2\,--\,4.
    
    As we study the CNM through HISA, we might miss a significant fraction of it. Since we can only trace gas that is cold enough to be observed in absorption against a warmer background, we are limited in our HISA detection by the requirement of sufficient background emission. The CNM has temperatures $\lesssim 300\rm\,K$ \citep{1977ApJ...218..148M,2003ApJ...587..278W}. For future investigations, simulations could help to quantify the fraction of CNM that is invisible to our HISA method.
    The computed $\rm H_2$ column density and mass has an uncertainty of at least a factor of two due to the uncertainties in the value for the CO-to-$\rm H_2$ conversion \citep{2005ApJ...634.1126M,2012MNRAS.423.2342F,2014A&A...570A..65G}. Furthermore, we could miss a significant fraction of CO-dark $\rm H_2$ column density \citep{2008ApJ...679..481P,2009ApJ...692...91G,2013A&A...554A.103P}. Simulations suggest that the fraction of CO-dark $\rm H_2$ could even be as high as $\sim$50\% at conditions typical of the Milky Way disk \citep{2014MNRAS.441.1628S,2016MNRAS.458.3667D}. However, this fraction should be moderate in low-temperature environments \citep{2016MNRAS.462.3011G}.
\section{Discussion}
\subsection{Kinematics}\label{sec:discussion_kinematics}
    The histograms of line-of-sight peak velocities derived from HISA and \element[ ][13]{CO} generally agree for both parts of the filament. The results of the Gaussian fits to the spectra might not always reflect the actual kinematic structure as \element[ ][13]{CO} emission exhibits multiple velocity components in some regions between 37--$50\rm\,km\,s^{-1}$. However, the \element[ ][13]{CO} velocities generally agree with the HISA features in a statistical sense, even in the eastern part of the filament where we do not observe a spatial correlation along the line of sight. We take this as a confirmation that our extracted HISA features are in fact due to self-absorption.
    
    The line widths of HISA show a broad distribution of 2\,--\,$10\rm\,km\,s^{-1}$. The eastern part reveals enhanced HISA line widths, which are 3\,--\,$4\rm\,km\,s^{-1}$ higher toward the north of the molecular filament. Although speculative, this could be a signature of the compression of \ion{H}{i} gas passing through the spiral arm potential and triggering $\rm H_2$ formation \citep{2004ApJ...612..921B}. As the gas is leaving the spiral arm structure, this could inject turbulence on the downstream side that enhances the line widths. Although observationally difficult to distinguish, simulations of the galactic dynamics of the ISM suggest that there are systematic differences in velocity dispersion between molecular clouds within the spiral arm potential and inter-arm clouds \citep{2015MNRAS.447.3390D,2016MNRAS.458.3667D,2017MNRAS.470.4261D}. The morphological and kinematic differences in each part of the filament could therefore be related to its position with respect to the spiral arm potential. However, in order to differentiate between different scenarios, we need to investigate synthetic \ion{H}{i} observations, which is beyond the scope of our current analysis. We note that the broadened HISA lines toward some positions of the cloud might be subject to resolution effects and could be the superposition of multiple lines. Spectrum~5 in Fig.~\ref{fig:baseline_fits} clearly shows multiple \element[][13]{CO} components where we detect an enhanced HISA line width. The lack of spatial correlation between HISA and \element[][13]{CO}, particularly in the eastern region, makes it difficult to assess if multiple \element[][13]{CO} components are preferentially associated with enhanced HISA line widths.
    Since the velocity dispersion is mostly due to turbulence in both tracers, we conclude that the agreement in velocities is robust.
\subsection{Column density probability density functions (N-PDFs)}\label{sec:N-PDFs}
    The column density maps derived in Fig.~\ref{fig:column_density_map} can be further evaluated by determining their probability density function (PDF). Column or volume density PDFs are commonly used as a measure of the density structure and physical processes acting within the cloud \citep[e.g.,][]{2014Sci...344..183K}.
    A log-normal shape of the N-PDF is usually attributed to turbulent motions dominating the early diffuse phase of a cloud's evolution.
    Furthermore, the width of the log-normal distribution reflects the amplitude of turbulence and can be associated with the Mach number \citep[e.g.,][]{1997MNRAS.288..145P,1998PhRvE..58.4501P,2007ApJ...665..416K,2008PhST..132a4025F}. In later evolutionary stages, molecular clouds can develop high-density regions due to the increasing effect of self gravity, producing a power-law tail in their N-PDF. Molecular cloud complexes that show star-forming activity favor this scenario as they reveal such power-law tails \citep{2009A&A...508L..35K,2013ApJ...766L..17S,2016A&A...587A..74S}.
    
    The shape of the resulting N-PDF is also sensitive to the regions where column densities are taken into account, especially in the low column density regime \citep{2015A&A...576L...1L}, and it is sensitive to the treatment of zero spacing information in interferometric data \citep{2016A&A...590A.104O}.
    We derived each N-PDF from the regions marked by the red polygons in Fig.~\ref{fig:column_density_map}, respectively. However, if we take into account low column densities that extend beyond the region enclosed by the polygons, we will miss a significant fraction of low column densities and hence the shape of the N-PDF will not recover the structure at the lower end well. We therefore chose to derive the N-PDFs from column densities approximately within the last closed contours that are still within the selected polygon regions. In order to compare the \ion{H}{i} column densities with those of molecular hydrogen, we converted $N(\rm H_2)$ to $N(\rm H)$ to construct the $\rm H_2$ N-PDFs. For the N-PDFs, we chose closed contours of $1.9\times 10^{20}$ and $3.2\times 10^{21}\rm\,cm^{-2}$ for HISA and $\rm H_2$, respectively. The selected closed contours that go beyond the polygon regions do not significantly change the shape of the N-PDFs. We set the column density threshold for \ion{H}{i} emission to $2.7\times 10^{21}\rm\,cm^{-2}$. It is difficult to define a last closed contour for \ion{H}{i} emission. However, this boundary bias has a negligible impact on the shape of the \ion{H}{i} N-PDF as we observe a small range of column densities due to the diffuse nature of \ion{H}{i} emission. For turbulence-dominated gas, last closed contours are not essential to sample the N-PDF properly \citep{2019MNRAS.482.5233K}. We furthermore normalized each N-PDF by the mean column density.
    \begin{figure*}[!htbp]
      \centering
        \includegraphics[width=1.0\textwidth]{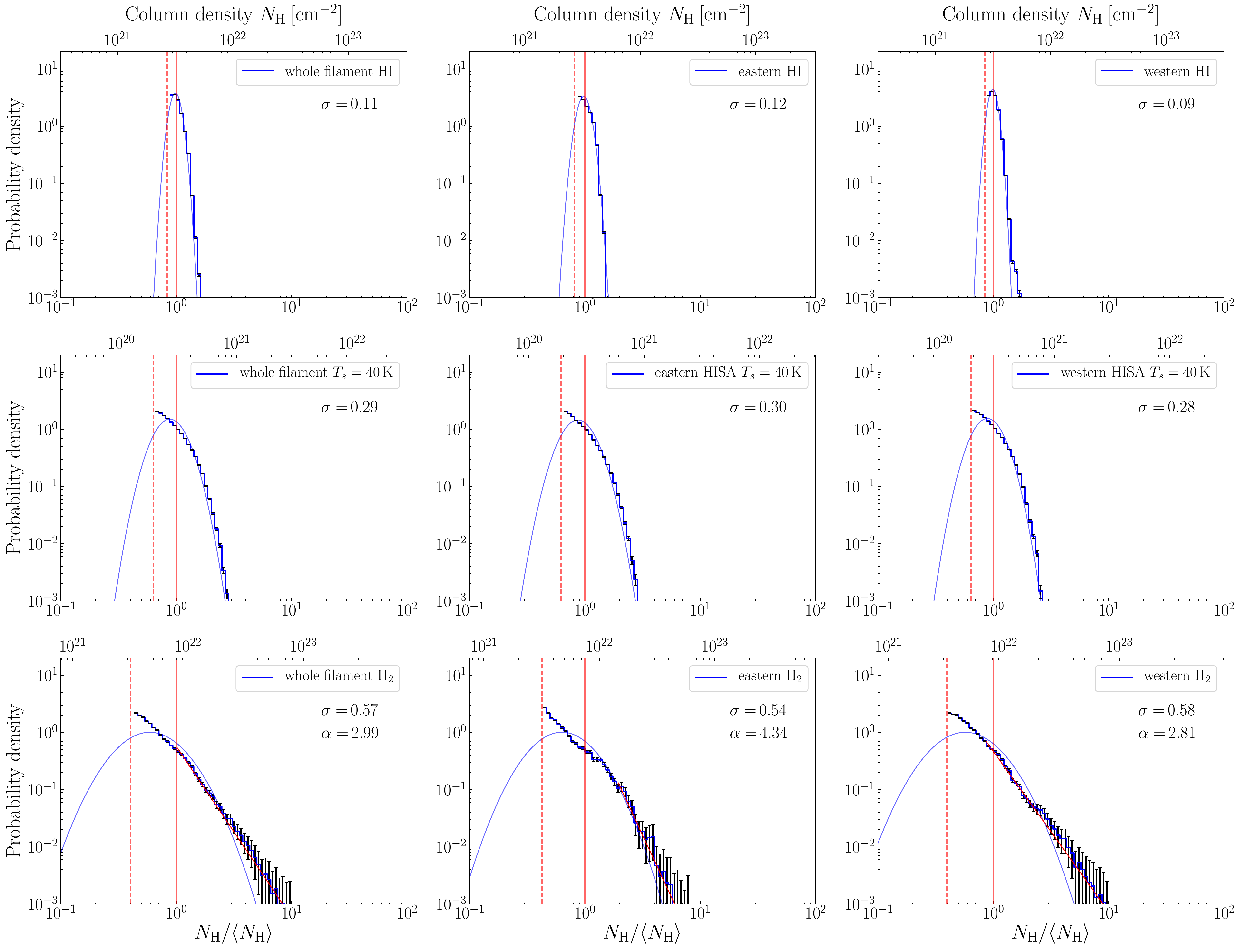}
      \caption[]{\textit{Top panels:} N-PDFs traced by \ion{H}{i} emission. The distributions are derived from the \ion{H}{i} column densities that have been corrected for optical depth and continuum emission (top panel of Fig.~\ref{fig:column_density_map}). \textit{Middle panels:} The N-PDFs of the gas traced by HISA. \textit{Bottom panels:} $\rm H_2$ N-PDFs traced by \element[ ][13]{CO}. The left panels show the derived N-PDF of the whole filament (east+west), respectively. The middle and right panel show the N-PDFs of the eastern and western part of the filament, respectively. The blue curves indicate the log-normal fits to the distribution. The red vertical dashed and solid lines mark the column density threshold (last closed contour) and mean column density, respectively. The red solid lines in the lower panels indicate the fit to the power-law tail.}
      \label{fig:column_density_PDFs}
    \end{figure*}
    Figure~\ref{fig:column_density_PDFs} presents the N-PDFs of \ion{H}{i} emission, HISA, and $\rm H_2$ column densities for each part of the filament (east/west) as well as the whole filament (east+west), respectively.
    \begin{table}[!htbp]
        \caption{Results of the fits to the N-PDFs.}
        \renewcommand*{\arraystretch}{1.3}
        \centering
        \begin{tabular}{c c c c}
             \hline\hline
             Component & $\langle N_{\mathrm{H}}\rangle$ [$\rm cm^{-2}$] & Width $\sigma$ & PL index $\alpha$ \\\hline
             \ion{H}{i} (WNM+CNM) & & & \\
             \hline
             Whole filament & $3.2\times 10^{21}$ & 0.11 & - \\
             East & $3.3\times 10^{21}$ & 0.12 & - \\
             West & $3.2\times 10^{21}$ & 0.09 & - \\\hline
             HISA (CNM) & & & \\
             \hline
             Whole filament & $3.0\times 10^{20}$ & 0.29 & - \\
             East & $3.0\times 10^{20}$ & 0.30 & - \\
             West & $3.0\times 10^{20}$ & 0.28 & - \\\hline
             $\rm H_2$ & & & \\
             \hline
             Whole filament & $7.6\times 10^{21}$ & 0.57 & 2.99 \\
             East & $7.3\times 10^{21}$ & 0.54 & 4.34 \\
             West & $7.7\times 10^{21}$ & 0.58 & 2.81 \\\hline
             All gas & & & \\
             \hline
             Whole filament & $8.3\times 10^{21}$ & - & 3.53 \\
        \end{tabular}
        \tablefoot{
        The second column shows the mean column density of each component designated in the first column. The third column presents the widths of the log-normal function fitted to the N-PDFs. The last column shows the index of the power-law (PL) function fitted to the tail of the $\rm H_2$ and All gas N-PDFs.
        }
        \label{tab:PDF_fitting}
    \end{table}{}
    
    As expected from the column density maps in Fig.~\ref{fig:column_density_map}, the N-PDF of the CNM as traced by HISA peaks at lower column densities than molecular hydrogen. The HISA N-PDFs are well represented by a log-normal function. The results of log-normal fits are shown in Table \ref{tab:PDF_fitting}. The log-normal shape implies that turbulent motions might be dominant and gravitational collapse leading to high column density peaks is not visible in HISA within the whole filament. There is no significant difference between the subregions defined in the eastern and western part of the filament. The widths of the HISA N-PDFs are the same for both regions. The mean column density derived from HISA is $\sim$3$\times 10^{20}\rm\,cm^{-2}$.
    
    The mean column densities and widths of the N-PDFs agree well with those found by \citet{2020A&A...634A.139W} for \mbox{GMF38.1-32.4a.} To investigate how observational uncertainties affect the width of the N-PDF, \citet{2020A&A...634A.139W} created model images of \ion{H}{i} and continuum emission with similar properties and noise as the real THOR data. They introduced artificial \ion{H}{i} absorption features from known column density distributions and added them to the model data. They extracted the HISA features using a similar method and derived column density distributions showing that the widths of the N-PDFs they find do not significantly increase due to observational uncertainties or the HISA extraction method. They conclude that the widths of the derived N-PDFs are robust and not subject to broadening introduced by observational noise and the fitting approach.
    
    The mean column densities of molecular hydrogen are about an order of magnitude higher than the column densities of HISA. In contrast to the HISA N-PDFs, the N-PDFs of molecular hydrogen are poorly represented by a log-normal function. Even though the eastern region does not show similarly high column density peaks as the western region, a power-law behavior is evident in both column density distributions. Therefore, power-law functions ($p(x)\propto x^{-\alpha}$) were additionally fitted to the high column density tail of the $\rm H_2$ N-PDFs. The best minimal column density for the power-law fit is obtained from the minimal Kolmogorov-Smirnov distance between the fit and the N-PDF. The fits were performed using the python package \mbox{\textit{Powerlaw}} \citep{10.1371/journal.pone.0085777}. The fitted parameters of the power-law functions are also listed in Table \ref{tab:PDF_fitting}.
    
    Power-law tails can be a sign of gravitational collapse, which creates high column density peaks \citep{2000ApJ...535..869K,2008PhST..132a4025F,2009A&A...508L..35K,2016A&A...587A..74S}. In agreement with observations, simulations of self-gravitating, turbulent molecular clouds show that star-forming activity reveals strong deviations from the log-normal shape in the form of power-law tails toward high column densities \citep{2011ApJ...727L..20K}. The slope of the power-law tails can then be associated with the evolutionary stage of the cloud, with shallower slopes indicative of an increasing degree of star formation efficiency \citep{2013ApJ...763...51F}. In general, theoretical studies and simulations of molecular clouds can reproduce N-PDFs of different forms, depending on the degree of turbulence, star-forming activity, and magnetic field support \citep{1994ApJ...423..681V,2010A&A...512A..81F,2015ApJ...808...48B}. Both subregions miss a small fraction of low column densities above the closed contour threshold. However, the shape of the $\rm H_2$ N-PDFs does not change significantly if we take into account all closed contours beyond the polygon regions.
    
    The N-PDFs derived from the \ion{H}{i} emission that traces a combination of CNM and WNM show a narrow log-normal shape with widths of $\sigma=0.09 - 0.12$. Observations toward well-known molecular cloud complexes also show N-PDFs of \ion{H}{i} emission with narrow log-normal shapes \citep{2015ApJ...811L..28B,2016ApJ...829..102I,2017MNRAS.472.1685R}. We might overestimate the column densities as the optical depth derived from absorption (see Sect.~\ref{sec:HI_emission_optical_depth}) is mostly due to cold atomic gas acting as the absorbing medium. However, we used this optical depth measurement to correct for \ion{H}{i} emission that might also be attributed to warm and optically thin gas. \citet{2015A&A...580A.112B} assess this effect and investigate the overestimation by comparing the corrected total \ion{H}{i} column densities with actual column densities of known spin temperatures and optical depths. They show that this overestimate is $<10\%$ for measured CNM optical depths $\tau<1.5$. This effect is therefore negligible compared to the uncertainty of the optical depth measurement itself. Furthermore, this systematic effect does not significantly affect the shape of the N-PDF.
    
    The mean column densities inferred from \ion{H}{i} emission are $\langle N_{\rm HI}\rangle\sim 3\,\times 10^{21}\rm\,cm^{-2}$. The \ion{H}{i} column densities show a narrow log-normal distribution driven by turbulent motions whereas the N-PDFs of molecular hydrogen show a broad distribution with a power-law behavior toward high column densities that might be subject to gravitational collapse. The column densities traced by \ion{H}{i} emission are an order of magnitude higher than those traced by self-absorption. The narrow width of the N-PDF represents the diffuse nature of \ion{H}{i} emission while the broader column density distribution traced by HISA indicates a clumpier structure of the CNM.
    
\begin{figure}[!htbp]
      \centering
        \resizebox{\hsize}{!}{\includegraphics{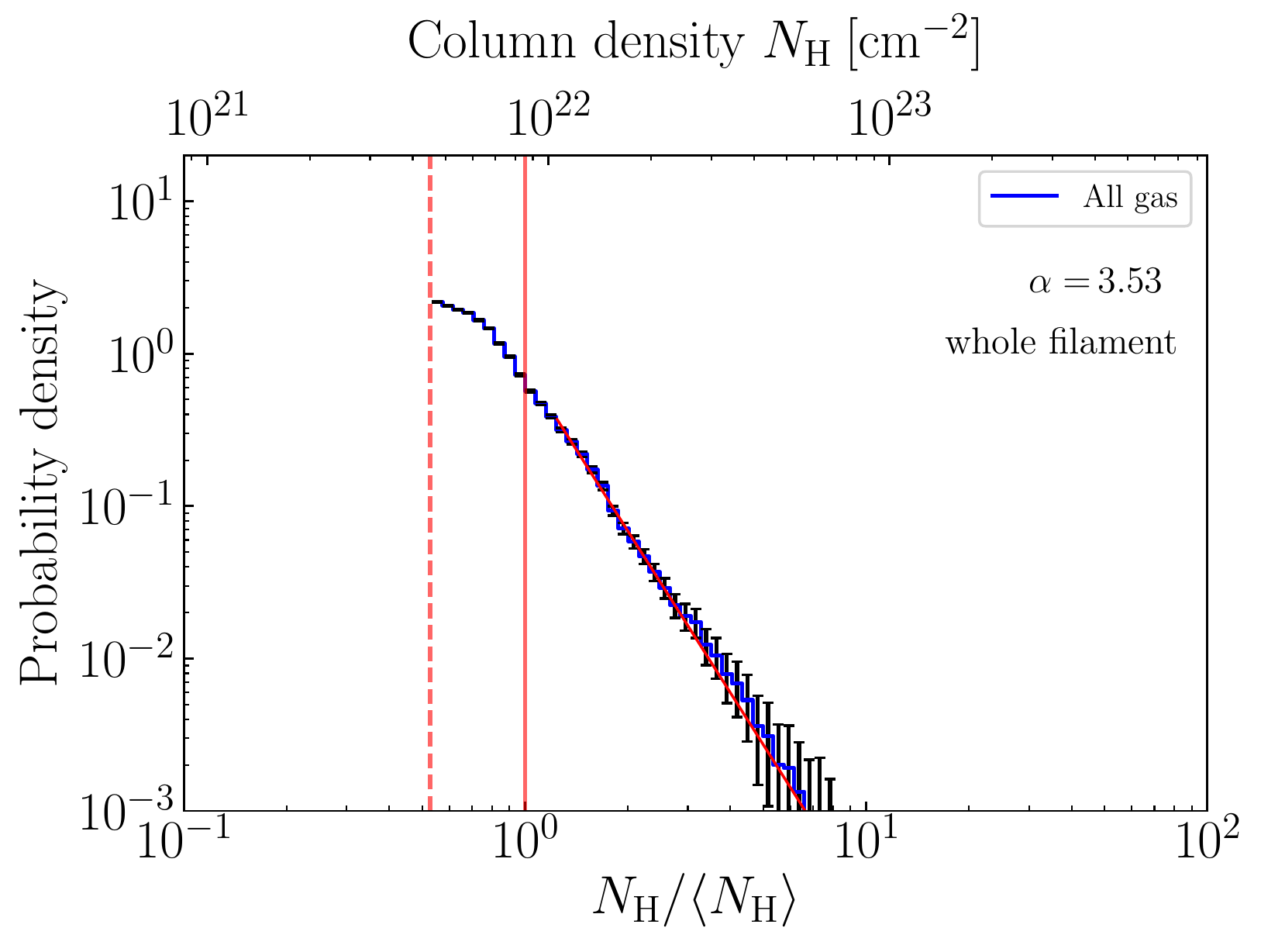}}
      \caption[]{All gas N-PDF of GMF20.0-17.9. The PDF is derived by adding the column densities of \ion{H}{i}, HISA, and $\rm H_2$. The plot shows the derived N-PDF of the whole filament marked by both the red polygons in Fig.~\ref{fig:column_density_map}. The red vertical dashed and solid line marks the column density threshold (last closed contour) at $4.5\times10^{21}\rm\,cm^{-2}$ and mean column density, respectively. The red solid line indicates the power-law fit to the high column density tail of the distribution.}
      \label{fig:all_gas_PDF}
    \end{figure}

    We examined the column density distribution of the entire hydrogen content of the filament, that is, both the atomic and molecular phase of GMF20.0-17.9. We derived an ``All gas'' N-PDF in Fig.~\ref{fig:all_gas_PDF} by adding together the column densities of all three tracers.
    We fitted the high column density tail of the distribution with a power-law function. The N-PDF can be very well described by a single power-law function.

    The western part shows higher $\rm H_2$ column density peaks and a shallower power-law tail in the N-PDF. The $\rm H_2$ column densities are generally lower in the eastern part of the filament. The ATLASGAL survey \citep{2009A&A...504..415S} reveals several high-density clumps in the western part of the filament and few clumps in the eastern part within the \element[][13]{CO} velocity range.
    The N-PDFs traced by HISA do not show significant differences between each part of the filament. However, the mass of HISA compared to its molecular counterpart does therefore increase toward the eastern part of the filament. The maximum spin temperature of our extracted HISA features is also lower in the eastern subregion (Fig.~\ref{fig:HISA_max_spin_temperature}). This might be an indication that the eastern subfilament is a young, cold \ion{H}{i} cloud while the western region exhibits a more evolved structure and star-forming activity. To furthermore test the validity of this hypothesis, we would need to extend our analysis to a larger sample of GMFs to deduce statistical evidence. This will be addressed in a future analysis. Simulations of cloud formation could also give constraints on signatures of kinematics and column densities in atomic and molecular line tracers. However, this is beyond the scope of this current investigation.
    
\subsection{Signatures of phase transition}
\begin{figure}[!htbp]
      \centering
        \resizebox{\hsize}{!}{\includegraphics{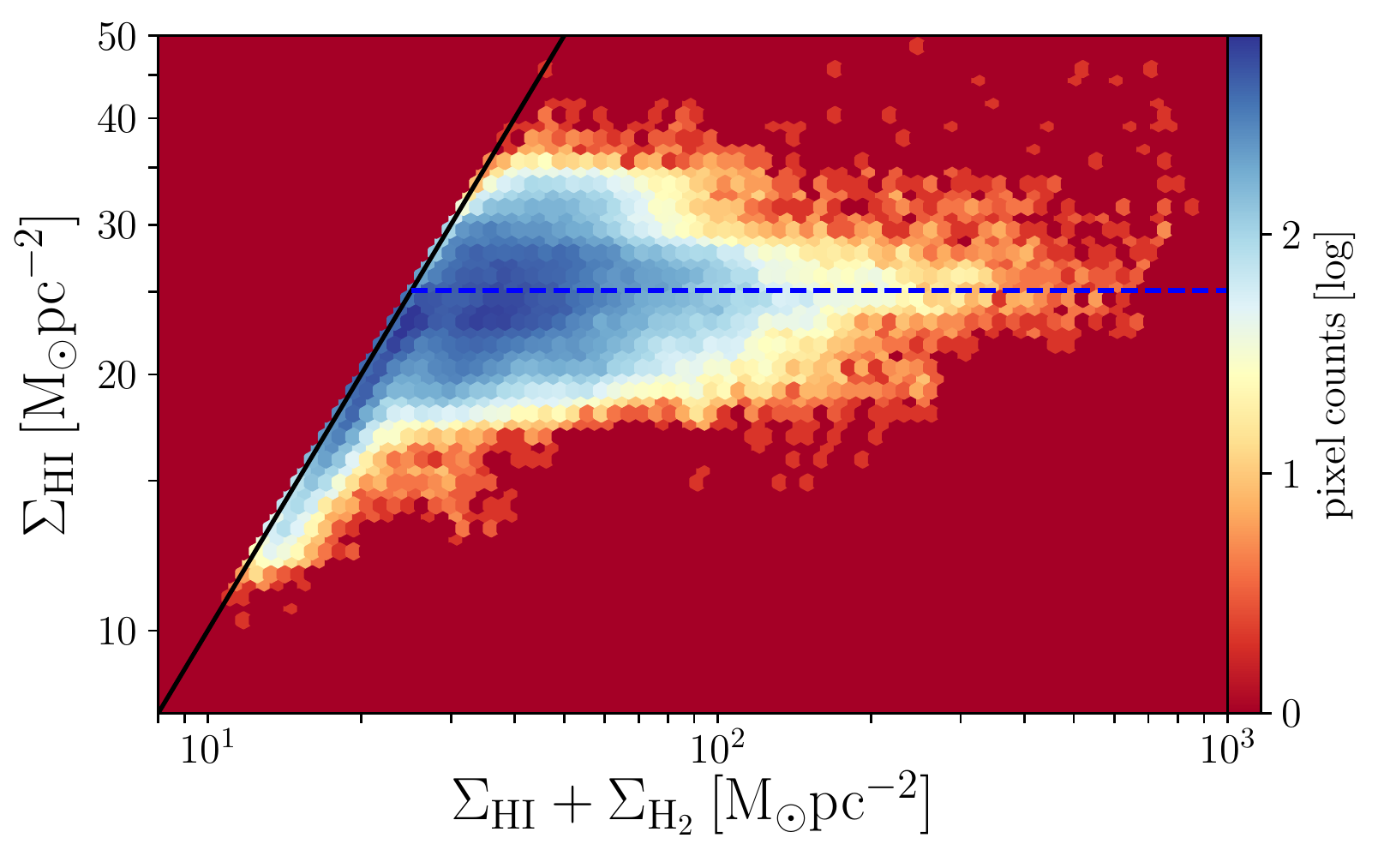}}
      \caption[]{\ion{H}{i}-to-$\rm H_2$ transition. The plot shows the \ion{H}{i} mass surface density traced by \ion{H}{i} emission as a function of the total hydrogen mass surface density. The black solid line indicates a 1-to-1 relation. The dashed blue line shows a fit to the mean of the distribution.}
      \label{fig:HI-to-H2_transition}
    \end{figure}

    The conversion of atomic to molecular gas (\ion{H}{i}-to-$\rm H_2$) is fundamental for molecular cloud formation processes. Theoretical models predict for a single \ion{H}{i}-to-$\rm H_2$ transition a mass surface density threshold of $\Sigma_{\rm HI}\sim5$\,--\,10$\rm\,M_{\odot}\,pc^{-2}$ for solar metallicity \citep{2008ApJ...689..865K,2009ApJ...693..216K,2014ApJ...790...10S}. In such models, the \ion{H}{i}-to-$\rm H_2$ transitions are computed assuming a balance between far-UV photodissociation and molecular formation, and accounting for the rapid attenuation of the radiation field due to $\rm H_2$ self-shielding and dust absorption \citep[see also][]{2016SAAS...43...85K}.
    Figure~\ref{fig:HI-to-H2_transition} presents the atomic hydrogen as a function of the total hydrogen mass surface density. We take into account all \ion{H}{i} column densities traced by the corrected \ion{H}{i} emission between $20.6>\ell>17.6^{\circ}$ and $-1.25<b<+0.5^{\circ}$. The figure reveals a saturation of atomic hydrogen at a mass surface density of $\sim$20\,--\,30$\rm\,M_{\odot}\,pc^{-2}$. A least squares fit to the mean of the distribution yields a mass surface density threshold of $\sim$25$\rm\,M_{\odot}\,pc^{-2}$ ($=3\times 10^{21}\rm\,cm^{-2}$). When examined individually, both the eastern and western subregion show the same \ion{H}{i} saturation level within the uncertainties ($24$ and $26\rm\,M_{\odot}\,pc^{-2}$, respectively). This transition exceeds the column density threshold predicted by theoretical models \citep{2008ApJ...689..865K,2009ApJ...693..216K,2014ApJ...790...10S}. \citet{2015A&A...580A.112B} report a column density threshold of 50\,--\,80$\rm\,M_{\odot}\,pc^{-2}$ toward the star-forming region W43, which is significantly higher than predicted transitions at $\sim$5\,--\,10$\rm\,M_{\odot}\,pc^{-2}$. \citet{2017ApJ...835..126B} argue that such high mass surface density thresholds cannot be explained by typical physical properties of the CNM as it would require an unrealistically high UV radiation field or low dust-to-gas ratio. As the clumpiness of a molecular cloud might regulate how far UV radiation penetrates the medium \citep[e.g.,][]{1988ApJ...332..379S,1991ApJ...374..522S}, \citet{2017ApJ...835..126B} suggest that the high thresholds can naturally be explained by a superposition of multiple transition layers observed along the line of sight. These authors predict a mass surface density threshold of $\sim$13$\rm\,M_{\odot}\,pc^{-2}$ for the more active star-forming region W43.
    \citet{2020A&A...634A.139W} find similar values of 14\,--\,23$\rm\,M_{\odot}\,pc^{-2}$ toward GMF38.1-32.4a, where the atomic gas surface density saturates to an almost flat distribution.
    
    While the derived atomic mass surface densities are a result of the combined column densities of WNM and CNM, the shielding from dissociating Lyman-Werner (LW) photons provided by a transition layer between atomic and molecular gas should be dominated by the CNM \citep{2009ApJ...693..216K}. The $\rm H_2$ formation rate per atom scales as the number density $n$, so the CNM, due to its higher density, is far more effective at shielding than the WNM. The observed transition should therefore be an upper limit and the actual critical surface density is $\leq 25\rm\,M_{\odot}\,pc^{-2}$, depending to the first approximation on the ratio $\Sigma_{\rm CNM}/\Sigma_{\rm WNM}$.
    Taking these considerations into account, we conclude that we observe at most 3--5 transition layers of atomic to molecular gas between 43 and $56\rm\,km\,s^{-1}$.
    
\subsection{Spatial correlation between atomic and molecular gas}\label{sec:HOG}
    The Histogram of Oriented Gradients\footnote{\url{https://github.com/solerjuan/astroHOG}} (HOG) is a method based on machine vision to study the spatial correlation in the emission by two or more spectral line tracers across velocity channels in an unbiased and systematic way. In Appendix~\ref{sec:appendix_HOG} we briefly outline the basic principles involved in this method. A comprehensive description is given by \citet{2019A&A...622A.166S}.
    
\begin{figure}[!htbp]
      \centering
        \resizebox{\hsize}{!}{\includegraphics{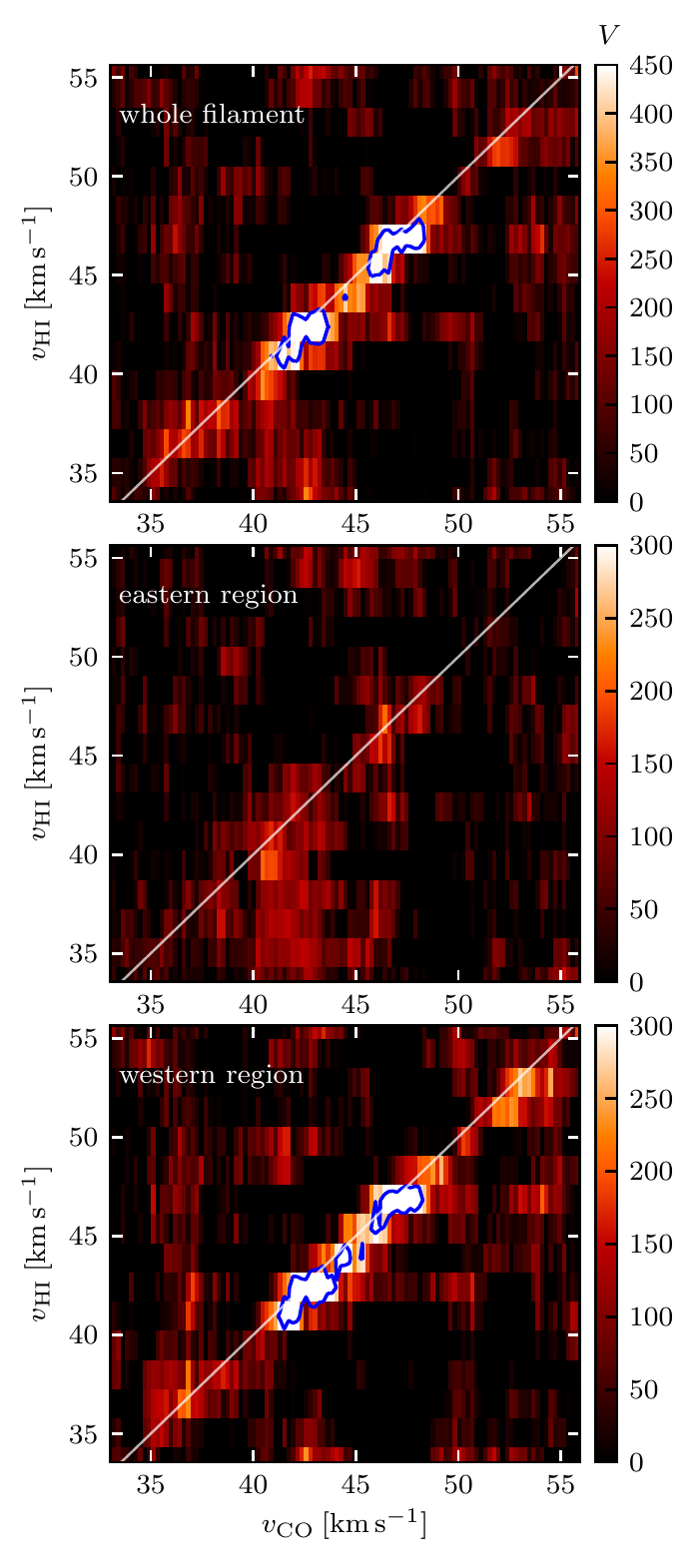}}
      \caption[]{HOG correlation plane of the filament as marked by the red polygons in Fig.~\ref{fig:column_density_map}. This figure presents the computed correlation between \ion{H}{i} and \element[ ][13]{CO} across velocities defined by the projected Rayleigh statistic. The white line shows a 1-to-1 correlation across velocities. The blue contours show the $5\sigma$ level on $V$. Large values of $V$ indicate a high spatial correlation. Values of $V$ close to zero indicate a negligible spatial correlation.}
      \label{fig:HOG_vplane_whole_filament}
    \end{figure}

    We applied the HOG on each part of the filament to investigate the spatial correlation between \ion{H}{i} and \element[ ][13]{CO}. The output of the HOG analysis is a matrix where the rows and columns correspond to the different velocity channels in each tracer, as shown in Fig.~\ref{fig:HOG_vplane_whole_filament}. The number in each matrix position corresponds to the projected Rayleigh statistic ($V$), which is an optimal estimator of the morphological correlation between the velocity channel maps as evaluated by the orientation of its intensity gradients.
    High values of $V$ correspond to high spatial correlation and values of $V$\,$\approx$\,0 correspond to very low spatial correlation.
    The intensity gradients are calculated using a Gaussian derivative kernel whose width determines the spatial scales under consideration. To exploit the available spatial resolution, we selected a derivative kernel size that matches the synthesized beam size of the GRS \element[][13]{CO} data, that is, 46\arcsec.
    The projected Rayleigh statistic is a measure of the significance of the spatial correlation, $V$\,$\approx$\,$\sqrt{2}$ is roughly the equivalent of a $1\sigma$ deviation from complete lack of correlation, which corresponds to a flat distribution in the angles between the intensity gradients. However, the significance of the result also has to be evaluated with respect to the chance correlation that may be present between the velocity channel maps. We use the standard deviation of $V$ in the velocity range between 10 and 90 km\,s$^{-1}$ as an estimate of the amplitude of the chance correlation against which we can evaluate the significance of the $V$ values. This assumes that there are enough independent velocity-channel maps in the selected velocity range.
    
    Figure~\ref{fig:HOG_vplane_whole_filament} presents the correlation distribution between \ion{H}{i} and \element[ ][13]{CO} for all parts of the filament as a function of velocity.
    We observe a significant spatial correlation in the velocity channels around $v_{\mathrm{HI}}\approx v_{^{13}\rm CO}\sim 43\rm\,km\,s^{-1}$ and $\sim47\rm\,km\,s^{-1}$ toward the west. However, the eastern part of the filament shows no significant correlation between \ion{H}{i} and \element[ ][13]{CO} emission at the velocities of GMF20.0-17.9. The observed correlation within the whole filament is therefore dominated by the western region.
    
    While we computed the spatial correlation between \ion{H}{i} and \element[][13]{CO} emission, we test the validity of the correlation by applying the HOG analysis to the inferred HISA and \element[][13]{CO} emission maps.
    The HOG yields similar findings for HISA and \element[][13]{CO}. As the absence of spatial correlation within the eastern part of the filament is reproduced in both analyses, we are confident that we do not observe any significant spatial correlation between \ion{H}{i} and \element[][13]{CO} within the eastern part of the filament.
    
    Small kernel sizes close or equal to the angular resolution of the telescope makes features produced by noise and nonideal telescope beams more evident. Since spatial correlation is expected across multiple scales \citep{1993MNRAS.262..327G,2000ApJ...537..720L,2001ApJ...555..130L}, we also examine the correlation in each analysis by setting the kernel size to $90\arcsec$, which is approximately twice the angular resolution of the THOR and GRS data ($40\arcsec$ and $46\arcsec$, respectively). The differences we find between the eastern and western region with our HISA method are reproduced by the HOG analysis, irrespective of the spatial scale we use. Thus, we consider these findings robust and not an artifact of our HISA extraction method.
    
    We conclude that the CNM appears to be associated with molecular gas in the western part whereas the molecular gas seems to be decoupled from its atomic counterpart in the more diffuse cloud envelope toward the east. The systematic differences in spatial correlation between east and west can be interpreted as an indication of different evolutionary stages.


\section{Conclusions}
    We have studied the atomic and molecular gas within the giant molecular filament GMF20.0-17.9. The molecular component is traced by GRS \element[ ][13]{CO} observations whereas the atomic gas is observed via \ion{H}{i} emission from the THOR survey. We isolated HISA features to disentangle the CNM from the atomic gas traced by \ion{H}{i} emission. We aimed to study the properties of the CNM as traced by HISA and compare our findings with the molecular counterpart. The results are summarized in the following:
    \begin{enumerate}
      \item We extracted HISA features by estimating the \ion{H}{i} emission spectrum in the absence of HISA. We employed a combination of first and second order polynomial functions to fit the baselines of HISA spectra at velocities adjacent to HISA features. This method gave the most reliable and robust results among the procedures we tested.
      \item The extracted HISA features reveal a spatial correlation with \element[ ][13]{CO} emission toward the western region of the filament while the eastern part shows no evidence that HISA traces the molecular gas. This finding is supported by the HOG analysis reporting significant spatial correlation toward the western part and no correlation toward the eastern part of the filament. However, the peak velocities of HISA and \element[ ][13]{CO} are in good agreement in both parts of the filament. The observed line widths of \element[ ][13]{CO} and HISA suggest that nonthermal effects like turbulent motions are the dominant driver for most regions within the filament.
      \item We derived $\rm H_2$ column densities from \element[ ][13]{CO} emission and compared the molecular column density distribution with its atomic counterpart. The HISA column densities show a more diffuse structure compared to those of molecular hydrogen. The $\rm H_2$ column densities reveal high-density peaks, particularly in the western part of the filament. The mass ratio of \ion{H}{i} (traced by HISA) and $\rm H_2$ is $0.01-0.05$, depending on the assumed spin temperature and region. This mass ratio increases toward the eastern part of the filament. The total \ion{H}{i} mass traced by \ion{H}{i} emission is similar to the molecular mass within the defined regions. The mass surface density threshold from the total \ion{H}{i} to $\rm H_2$ is observed to be $\sim$25$\rm\,M_{\odot}\,pc^{-2}$, in excess of predictions by theoretical models. However, this result can naturally be explained by a superposition of multiple transition layers or an additional WNM fraction that is less effective at shielding.
      \item The HISA N-PDFs can be well described by log-normal functions in both parts of the filament, indicative of turbulent motions as the main driver for these structures. While the magnitude of the column densities are dependent on the assumed parameters of spin temperature and background fraction, the shape and width of the N-PDFs are robust. The N-PDFs of \ion{H}{i} emission tracing both the WNM and CNM of the atomic gas represent the diffuse structure and show a narrow log-normal shape. The $\rm H_2$ column densities show a broad log-normal distribution with an indication of a power-law tail, more pronounced in the western part of the filament.
      \item We speculate that the two parts of the filament reflect different evolutionary stages. Interestingly, the derived HISA features in the eastern part of the cloud show lower maximum spin temperatures. This favors the scenario of a younger, less evolved cloud that is forming molecular gas out of the atomic gas reservoir. The western region harbors signs of active star formation and shows more pronounced column density peaks of $\rm H_2$. Moreover, the mass fraction of $\rm H_2$ compared to cold atomic hydrogen traced by HISA is larger toward the western part of the filament. While the HISA features correlate well with the molecular gas in the western part of the filament, they lack spatial correlation with the molecular component in the eastern region. Furthermore, we speculate that signatures of spiral arm interaction with atomic gas are visible toward the eastern part of the filament, due to an enhancement of line widths. The spatial structure and kinematics provide useful observables for theoretical models and simulations.
    \end{enumerate}
    A statistical treatment of the HISA properties in the Galactic plane is still missing. However, this case study toward a known large-scale filament, which is complementary to the analysis by \citet{2020A&A...634A.139W}, serves as a good laboratory to study the properties of the CNM.
\begin{acknowledgements}
    J.S., H.B., R.S.K., and S.C.O.G. acknowledge support from the Deutsche Forschungsgemeinschaft (DFG, German Research Foundation) -- Project-ID 138713538 -- SFB 881 (``The Milky Way System'', subprojects A01, B01, B02, and B08). Y.W., H.B., and J.D.S. additionally acknowledge support from the European Research Council under the Horizon 2020 Framework Program via the ERC Consolidator Grant CSF-648505. 
    R.S.K. and S.C.O.G. also thank for funding from the Heidelberg Cluster of Excellence \mbox{STRUCTURES} in the framework of Germany’s Excellence Strategy (grant EXC-2181/1 - 390900948). R.S.K. also thanks for funding from the European Research Council via the ERC Synergy Grant ECOGAL (grant 855130).
    F.B. acknowledges funding from the European Research Council (ERC) under the European Union’s Horizon 2020 research and innovation programme (grant agreement No.726384/Empire).
    This work was carried out in part at the Jet Propulsion Laboratory which is operated for NASA by the \mbox{California Institute of Technology}. This research made use of Astropy and affiliated packages, a community-developed core Python package for Astronomy \citep{2018AJ....156..123A}, Python package SciPy\footnote{\url{https://www.scipy.org}}, and APLpy, an open-source plotting package for Python \citep{aplpy2012}.
    The authors thank the anonymous referee for the detailed comments and constructive suggestions that improved the paper.
\end{acknowledgements}
%
%
\bibliographystyle{aa_url} 

\bibliography{references} 

\begin{appendix}
\section{\ion{H}{i} self-absorption uncertainties}\label{sec:discussion_extraction_method}
    To extract HISA features, we have employed a combination of first and second order polynomial functions to fit the HISA baselines at smoothed \ion{H}{i} spectra to estimate the \ion{H}{i} spectra in the absence of self-absorption ($T_{\mathrm{off}}$). Different polynomial functions have partly shown significant differences in the derivation of the background spectrum. In reality we do not know what the background spectrum would look like in the absence of self-absorption and it is therefore difficult to quantify the uncertainty of $T_{\mathrm{off}}$. We can estimate the systematic uncertainty of $T_{\mathrm{off}}$ due to our baseline fitting routine by the deviation of different polynomial function fits. We compare the baseline fits inferred from first and second order polynomial functions and calculate the standard deviation of the difference of the baseline fits. The difference has a mean of -2$\rm\,K$, indicating that the second order baseline fits on average overestimate the background with respect to first order fits. The standard deviation of the difference is $\sim$7$\rm\,K$. If we add the observational noise and fitting uncertainty of $8\rm\,K$ to the systematic error, the total uncertainty is $\Delta T_{\mathrm{off}}=11\rm\,K$. For the mean temperatures of $T_{\mathrm{on}}=40\rm\,K$ and $T_{\mathrm{off}}=70\rm\,K$ that we measure with our extraction method between 43--$56\rm\,km\,s^{-1}$, this propagates to a relative uncertainty of $\sim$40\% for the optical depth and therefore column density. While this uncertainty affects the amplitude of the HISA features, the centroid velocities should be robust and weakly dependent on the employed fitting method.
    
    The column densities and masses are furthermore affected by the integration over velocities. We integrated all derived column densities from 43 to $56\rm\,km\,s^{-1}$ as we extracted significant HISA features in that velocity regime. \citet{2014A&A...568A..73R} assign velocities between 37 and $50\rm\,km\,s^{-1}$ to the giant molecular filament as they define their filaments through smooth velocity gradients. Furthermore, velocities $\gtrsim 50\rm\,km\,s^{-1}$ might not be associated with the near Scutum-Centaurus spiral arm based on the model by \citet{2008AJ....135.1301V}. If we take into account velocities of 43--$51\rm\,km\,^{-1}$ for the derivation of the column densities, the column densities decrease by $\sim$30\%. However, employing the additional maser parallax information in the Bayesian distance calculator of the BeSSeL survey \citep{2014ApJ...783..130R,2016ApJ...823...77R,2019ApJ...885..131R} yields velocities up to $\sim$ 55$\rm\,km\,s^{-1}$ associated with the near Scutum-Centaurus spiral arm. We therefore included the velocities $>50\rm\,km\,s^{-1}$ for the column density calculation.
\subsection{Background fraction}\label{sec:discussion_background_fraction}
\begin{figure}[!htbp]
      \centering
        \resizebox{\hsize}{!}{\includegraphics{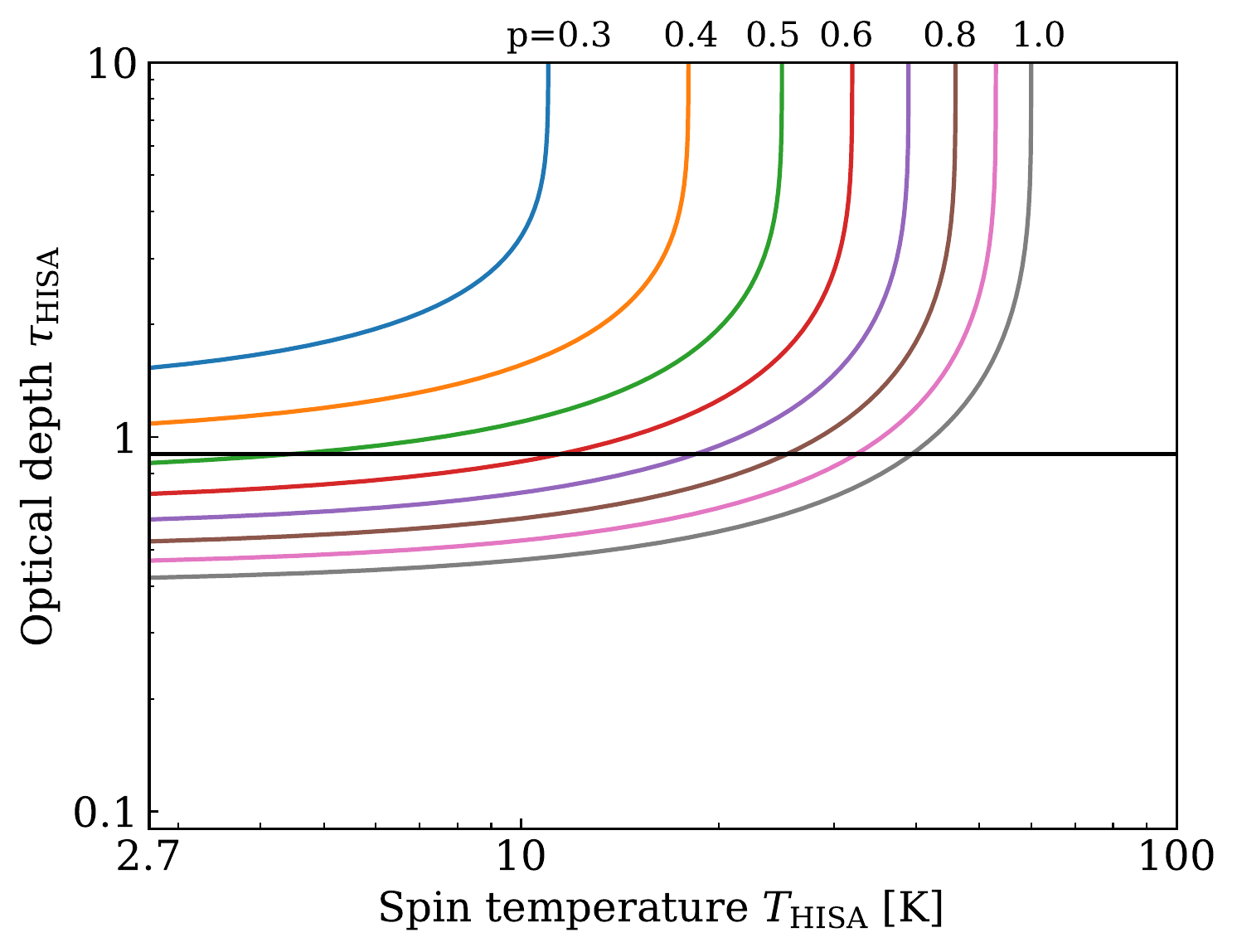}}
      \caption[]{Optical depth $\tau_{\mathrm{HISA}}$ as a function of spin temperature $T_{\mathrm{HISA}}$ at the derived mean temperatures of $T_{\mathrm{on}}=40\,\rm K$, $T_{\mathrm{off}}=70\,\rm K$, and $T_{\mathrm{cont}}=20\,\rm K$ (Eq.~\ref{equ:T_ON-T_OFF}). The different colors represent varying background fractions $p$, with values from $0.3$ to $1.0$. The horizontal black line marks an optical depth of 0.9 that was determined from \ion{H}{i} absorption against strong continuum emission.}
      \label{fig:tau_HISA_discussion}
    \end{figure}

    The HISA column density computation relies on assumptions of the spin temperature $T_{\mathrm{HISA}}$ and background fraction $p$. We assumed a background fraction of $p=0.9$ for the whole filament since we expect large contributions to the \ion{H}{i} emission originating in the background due to the distance and location of the filament as well as the kinematic distance ambiguity. Figure~\ref{fig:tau_HISA_discussion} shows the optical depth as a function of the spin temperature for different background fractions (Eq.~\ref{equ:T_ON-T_OFF}). Mean temperatures of $T_{\mathrm{on}}=40\rm\,K$, $T_{\mathrm{off}}=70\rm\,K$, and $T_{\mathrm{cont}}=20\rm\,K$ were adopted from our results. For a constant spin temperature the optical depth increases with decreasing background fraction $p$. Consequently, the column density increases with lower background fraction.
    To investigate the variation of the column density as a function of $p$, we calculated the HISA N-PDFs assuming background fractions of $p=0.7$, 0.8, and 0.9, respectively. Furthermore, we assumed a constant spin temperature of $T_{\mathrm{HISA}}=40\rm\,K$.
    \begin{figure}[!htbp]
      \centering
        \resizebox{\hsize}{!}{\includegraphics{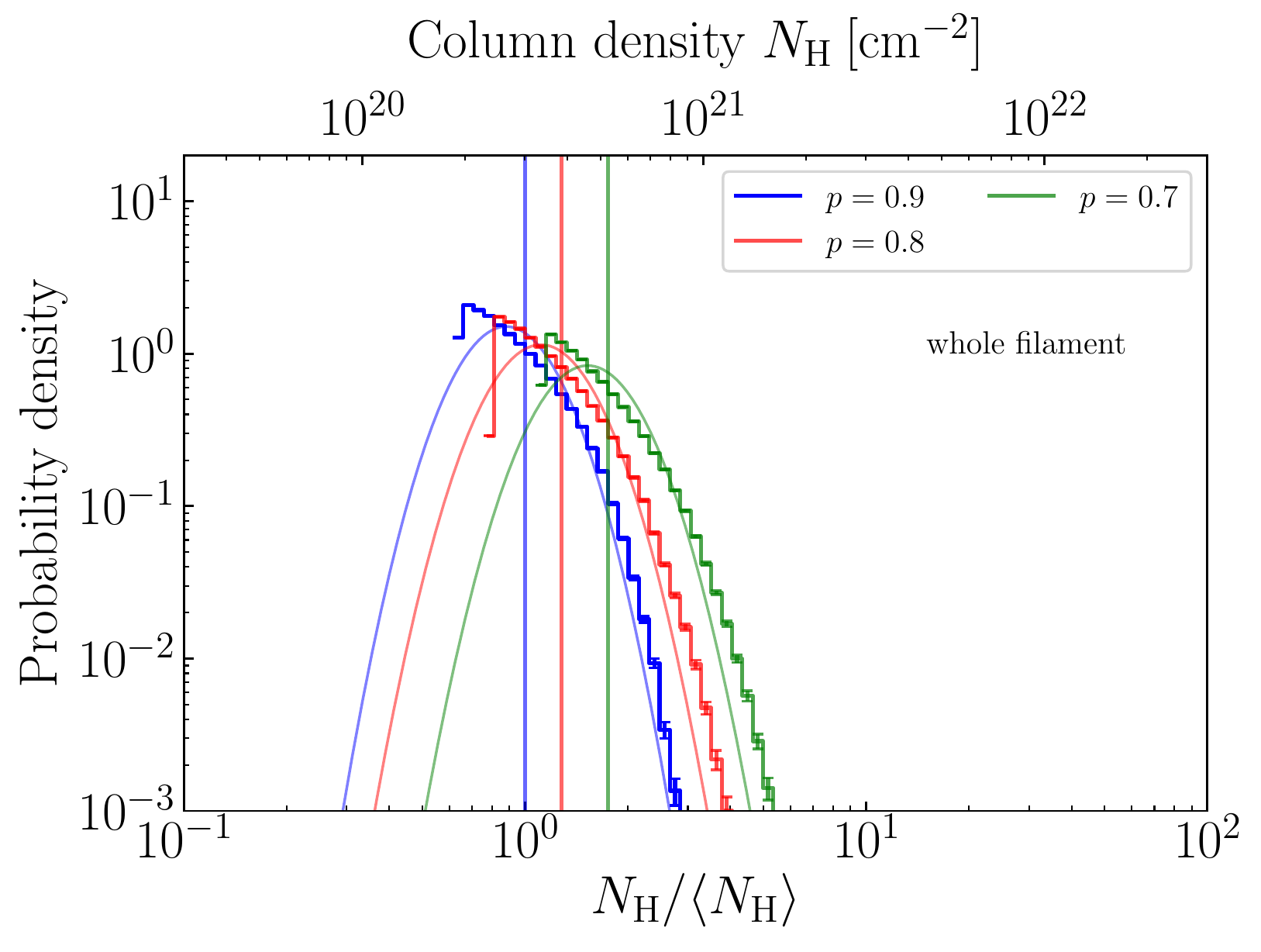}}
      \caption[]{\ion{H}{i} self-absorption N-PDFs for different values of background fraction $p$. The plot shows the derived N-PDF of the whole filament marked by both the red polygons in Fig.~\ref{fig:column_density_map}. The blue, red, and green distributions correspond to a background fraction of $0.9$, $0.8$, and $0.7$, respectively. The vertical lines show the corresponding mean column densities.}
      \label{fig:HISA_PDF_p_values}
    \end{figure}
    Figure~\ref{fig:HISA_PDF_p_values} shows the HISA N-PDFs for the whole filament. The shape and width of the N-PDF does not change with varying background fraction $p$. However, the mean column density increases by a factor of $\sim$2 from $p=0.9$ to $0.7$.
    We can furthermore estimate the amount of background emission from the radial \ion{H}{i} volume density distribution in the Galactic plane. For Galactocentric radii $7\lesssim R\lesssim 35\rm\,kpc$ \citet{2008A&A...487..951K} report an average mid-plane volume density distribution of $n(R)\sim n_0\,e^{-(R-R_{\odot})/R_n}$ with $n_0=0.9\rm\,cm^{-3}$, $R_{\odot}=8.5\rm\,kpc$, and $R_n=3.15\rm\,kpc$. Assuming a constant volume density of $n(R<7\,\mathrm{kpc})=n(R=7\rm\,kpc)$, we can integrate the densities along the line of sight and estimate the amount of gas up to the distance of the filament (foreground) and beyond (background). We thus obtained a background fraction of $\sim$0.92 by integrating up to a Galactocentric distance of $35\rm\,kpc$. This relation gives the averaged distribution of the northern and southern Galactic plane and could hold systematic differences in some regions \citep{2009ARA&A..47...27K}. Since the foreground and background emission stems from the more diffuse \ion{H}{i} component, we do not expect large fluctuations on small scales. Hence, the assumption of a constant background fraction over the whole cloud is reasonable.
\subsection{Maximum spin temperature}\label{sec:discussion_max_spin_temperature}
    We have measured cold \ion{H}{i} in absorption against strong continuum sources to directly determine the optical depth. However, we cannot map the entire filament as this method requires continuum sources strong enough to induce absorption features. Interpolating the optical depth between locations of individual measurements is not trivial. We have estimated the uncertainty of the optical depth by a factor of two. The shape of the N-PDF is not affected by different optical depths if assumed constant for the whole filament. Within the uncertainty of the optical depth the mean column density varies by a factor of $\sim$1.5.
    In the velocity range of the HISA features the absorption spectra toward some sources are saturated and we can only report lower limits of the optical depth. Since the optical depths are high, we are confident that we in general do observe actual HISA instead of low \ion{H}{i} emission.
    We can only measure the optical depth of HISA together with its spin temperature and disentangling these quantities is difficult. We therefore assumed a constant spin temperature for the whole cloud. This is a poor assumption as the spin temperature is likely to vary within the filament. However, the optical depth measurements toward strong continuum sources alleviates this problem. The optical depth measurement toward G19.075-0.287 is more accurate as it is not saturated. The optical depth of $\tau=0.9$ is shown in Fig.~\ref{fig:tau_HISA_discussion} as a black horizontal line. Assuming $p=0.9$ and adopting the derived mean temperatures for $T_{\mathrm{on}}$, $T_{\mathrm{off}}$, and $T_{\mathrm{cont}}$ reveals a spin temperature of $\sim$32$\rm\,K$. This is slightly lower than our assumed spin temperature of $40\rm\,K$. This is again a poor assumption since we adopt a constant optical depth for the whole cloud. However, we are able to constrain an upper limit of the spin temperature in the limit of high optical depths.
    
    In general we can solely measure the optical depth of cold \ion{H}{i} as a function of the spin temperature (Eq.~\ref{equ:T_ON-T_OFF}). Figure~\ref{fig:tau_HISA_discussion} shows this relation. For high optical depths the spin temperature reaches a maximum. Solving Eq.~\eqref{equ:T_ON-T_OFF} for $T_{\mathrm{HISA}}$ illustrates that
    
    \begin{equation}
    T_{\mathrm{HISA}} = \frac{T_{\mathrm{on}}-T_{\mathrm{off}}}{1-e^{-\tau_{\mathrm{HISA}}}} + p\,T_{\mathrm{off}} + T_{\mathrm{cont}} \: .
    \end{equation}{}
    
    \noindent Since $T_{\mathrm{on}}-T_{\mathrm{off}}$ is always negative in the case of self-absorption, $T_{\mathrm{HISA}}$ reaches a maximum if $\frac{T_{\mathrm{on}}-T_{\mathrm{off}}}{1-e^{-\tau_{\mathrm{HISA}}}}$ becomes minimal. This is the case for $\tau\to\infty$. The maximum spin temperature therefore is
    
    \begin{equation}
    T_{\mathrm{HISA}}(\mathrm{max.}) = T_{\mathrm{on}} + T_{\mathrm{cont}} - (1-p)\,T_{\mathrm{off}} \: .
    \end{equation}{}
    
\begin{figure*}[!htbp]
      \centering
        \includegraphics[width=1.0\textwidth]{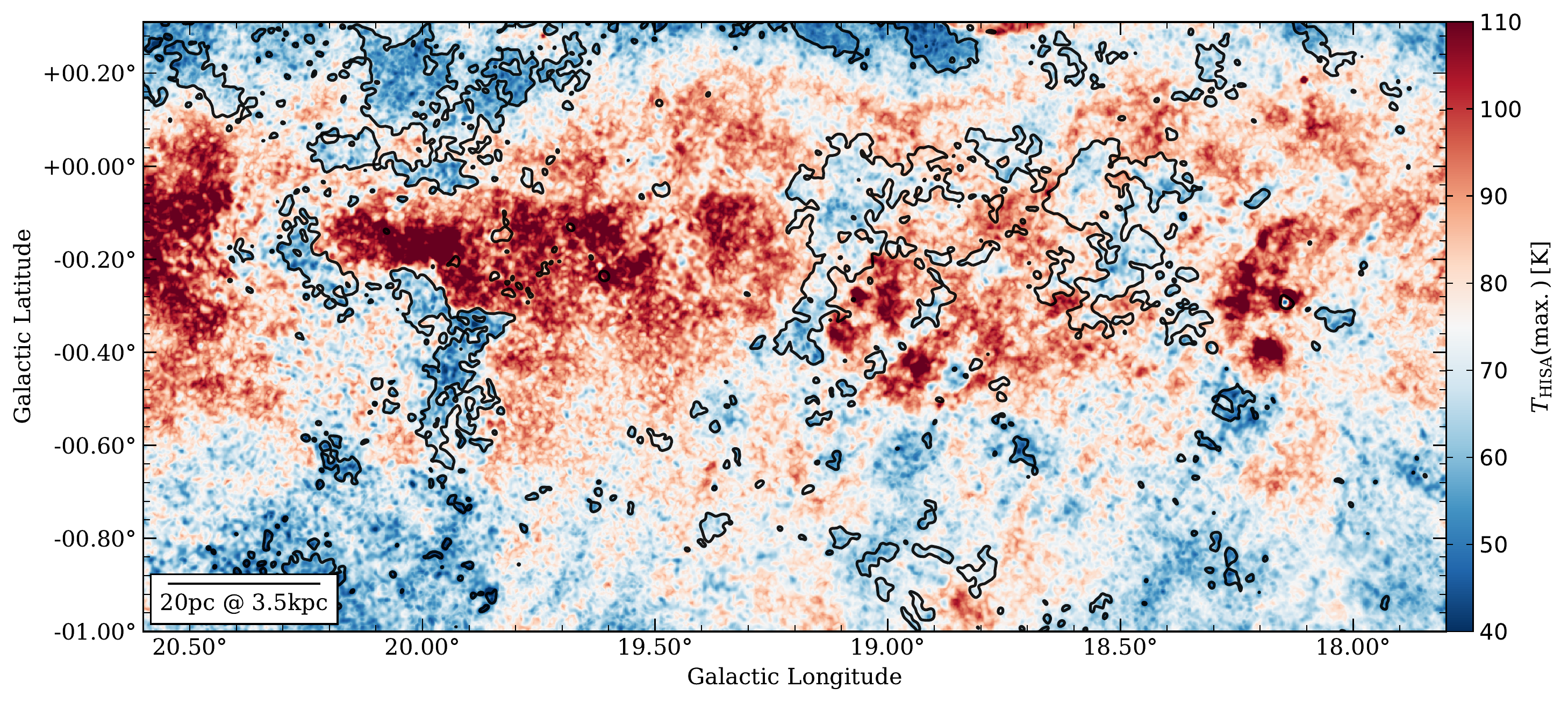}
      \caption[]{Maximum spin temperature of the HISA features assuming a background fraction of $p=0.9$. The black contour indicates the HISA column density at $3\times 10^{20}\rm\,cm^{-2}$.}
      \label{fig:HISA_max_spin_temperature}
    \end{figure*}

    \noindent The maximum spin temperature increases with increasing background fraction $p$. We use this relation to compute the upper limit of the spin temperature for each pixel in the map. We assumed a background fraction of $p=0.9$. Figure~\ref{fig:HISA_max_spin_temperature} presents the maximum spin temperature for the whole filament. Regions exposing continuum emission are not reliable as they contaminate HISA features. We see a clear correlation between the maximum spin temperature and the HISA column densities. As expected, we find the lowest maximum spin temperatures of $\sim$45$\rm\,K$ where we observe the strongest absorption features.
    For the derivation of the column densities and masses, we assumed a spin temperature of $40\rm\,K$. This is lower than the upper limit in any region of the filament. Therefore, the optical depths are not expected to be in the optically thick limit and the column densities we derived are reliable. Furthermore, we conclude that the estimated background fraction of $p=0.9$ at $T_{\mathrm{HISA}}=40\rm\,K$ is reasonable. However, as we can only report an upper limit of the spin temperature in the optically thick limit, we are not able to deduce the actual spin temperature of the HISA features. If we assume $p=0.7$, the maximum spin temperature drops by $\sim$20$\rm\,K$ and the assumed spin temperature of $40\rm\,K$ would exceed the upper limit for the coldest regions. The column density derivation of the coldest regions would therefore not be reliable anymore. In general, the spin temperature gives an additional constraint on the lower limit of the background fraction $p$ as the spin temperature would drop to unrealistically low values for certain background fractions.
\section{Histogram of oriented gradients}\label{sec:appendix_HOG}
    The Histogram of Oriented Gradients \citep[HOG;][]{2019A&A...622A.166S} is based on the assumption that the appearance and shape of an object can be characterized by the representation of its local intensity gradients. One simple application is deriving the spatial correlation between two images. If two images are spatially correlated, the relative angle $\varPhi$ between their local gradients is approximately $\Delta\varPhi\sim 0^\circ$ where the correlation is high. Two completely uncorrelated images would have a flat distribution of relative orientation angles $\varPhi$ while two identical images would disclose a Dirac delta distribution peaked at $\Delta\varPhi=0^\circ$ as all gradients are aligned.
    
    We used the HOG to analyze the spatial correlation between maps of \ion{H}{i} and \element[ ][13]{CO} across the radial velocities $v_{\rm LSR}$. We therefore regridded the \ion{H}{i} data cube to match the spatial grid and pixel size of the \element[ ][13]{CO} data for comparison. The gradients are computed using Gaussian derivatives. The derivatives are the result of the convolution of an image $f$ with the spatial derivative of a two-dimensional Gaussian function $\partial G^{(k)}$. The size of the Gaussian kernel $k$ determines the size over which the gradients are computed. Large kernel sizes compute gradients on large scales and therefore probe the correlation of the more diffuse structure of the cloud. Small scale correlation can be investigated by invoking small Gaussian kernel sizes. If the kernel size is too large, we smear out small-scale structures as we compute the average gradient over a larger spatial range.
    
    A quantity representing a measure for the spatial correlation between \ion{H}{i} at velocity $l$ and \element[ ][13]{CO} at velocity $m$ is the projected Rayleigh statistic ($V$). It is defined as
    
    \begin{equation}
    V_{lm} = \frac{\Sigma_{ij}\,w_{ij,lm}\,{\rm cos}(2\varPhi_{ij,lm})}{\sqrt{\Sigma_{ij}[(w_{ij,lm})^2/2]}} \: ,
    \label{equ:PRS}
    \end{equation}
    
    \noindent where the numerator is the sum over all pixels $i,j$ of the relative angles $\varPhi_{ij,lm}$ between the gradients of \ion{H}{i} and \element[ ][13]{CO} at the velocities $l$ and $m$, respectively. The statistical weight $w_{ij,lm}$ is introduced to account for spatial correlations of the gradients due to the telescope beam. The factor 2 in the cosine accounts for relative orientations between $[-\pi/2,+\pi/2]$, thus the correlation is measured independently of the direction of the gradients. Consequently, we are able to report both positive and negative (HISA) spatial correlation between \ion{H}{i} and \element[ ][13]{CO} emission.
    The projected Rayleigh statistic $V$ can be related to a random walk \citep[see e.g.,][]{2018MNRAS.474.1018J,2019ApJ...878..110F}, characterizing the distance from the origin after taking unit steps in the direction determined by each angle $\varPhi_{ij,lm}$. The expectation value of a random walk in two dimensions is zero, so any significant deviation from the origin would indicate preferential angles. We are particularly interested whether intensity gradients tend to be preferentially parallel ($\varPhi_{ij,lm}=0$) and how strong that preference is. The sum over a preferred orientation angle of $\varPhi_{ij,lm}=0$ will result in high $V$ values, just as a preferred orientation angle of $\varPhi_{ij,lm}=0$ in a "random" walk would result in a large distance from the starting point. Spatially uncorrelated data would reveal $V$ values of approximately zero.
    
    Associated physical structures probed by spectral line emission are not independent across velocities and inherently show a correlation as well. To assess the statistical significance of each velocity-channel map, we used the mean value $\langle V\rangle$ of the $V$ over the velocity range that we analyzed and estimated the standard deviation by the population variance of the distribution. We therefore assumed that over a broad velocity range the velocity channels are independent.
    Another uncertainty arises from observational noise in the data. The HOG addresses this by the use of Monte Carlo sampling to propagate the uncertainties in the observations. Therefore, for each velocity-channel map $n$ different realizations are generated within the observational uncertainty and with the same mean intensity. Using this sampling, the uncertainty of the correlation can be determined by estimating the variance of the correlation of different Monte Carlo realizations. Since we expect a contribution from nonuniform noise introduced by the observation, we report only $\geq 5\sigma$ confidence levels.
\end{appendix}
\end{document}